\documentclass[journal]{IEEEtran}

\usepackage{cite}      % Written by Donald Arseneau
\usepackage{graphicx,caption,subcaption}
\usepackage{amsmath}
\usepackage{amsfonts}
\usepackage{url}
\usepackage{color}
\usepackage[normalem]{ulem}

\newcommand{\be}{\begin{equation}}
\newcommand{\ee}{\end{equation}}
\newcommand{\bea}{\begin{eqnarray}}
\newcommand{\eea}{\end{eqnarray}}
\newcommand{\beaa}{\begin{eqnarray*}}
\newcommand{\eeaa}{\end{eqnarray*}}
\newcommand{\ben}{\begin{enumerate}}
\newcommand{\een}{\end{enumerate}}
\newcommand{\bi}{\begin{itemize}}
\newcommand{\ei}{\end{itemize}}

\newcommand{\lip}{\langle}
\newcommand{\rip}{\rangle}
\newcommand{\uu}{\underline}

\newcommand{\df}{{\rm d}}
\newcommand{\abs}[1]{\left\vert#1\right\vert}
\newcommand{\la}{\left(}
\newcommand{\ra}{\right)}

\begin{document}

\title{Pulse collision picture of {inter-channel} nonlinear interference in fiber-optic communications}

\author{Ronen Dar, Meir Feder, Antonio Mecozzi, and Mark Shtaif
%\thanks{Manuscript received \today}% <-this % stops a space
\thanks{R. Dar, M. Feder, and M. Shtaif are with with the School of Electrical Engineering,
Tel Aviv University, Tel Aviv 69978, Israel (emails: ronendar@post.tau.ac.il, meir@eng.tau.ac.il, shtaif@tauex.tau.ac.il). A. Mecozzi is with the Department of Physical and Chemical Sciences,
University of L'Aquila, L'Aquila 67100, Italy (email: antonio.mecozzi@univaq.it). }}

\maketitle

\begin{abstract}
We model the build-up of {inter-channel} nonlinear interference noise (NLIN) in wavelength division multiplexed systems by considering the pulse collision dynamics in the time domain. The fundamental interactions can be classified as two-pulse, three-pulse, or four-pulse collisions and they can be either complete, or incomplete. Each type of collision is shown to have its unique signature and the overall nature of NLIN is determined by the relative importance of the various classes of pulse collisions in a given WDM system. The pulse-collision picture provides qualitative and quantitative insight into the character of NLIN, offering a simple and intuitive explanation to all of the reported and previously unexplained phenomena. {In particular, we show that the most important contributions to NLIN follow from two-pulse and four-pulse collisions. While the contribution of two-pulse collisions is in the form of phase-noise and polarization-state-rotation with strong dependence on modulation format, four-pulse collisions generate complex circular noise whose variance is independent of modulation format. In addition, two-pulse collisions are strongest when the collision is complete, whereas four-pulse collisions are strongest when the collision is incomplete. We show that two-pulse collisions dominate the formation of NLIN in short links with lumped amplification, or in links with distributed amplification extending over arbitrary length. In long links using lumped amplification the relative significance of four-pulse collisions increases, emphasizing the circularity of the NLIN while reducing its dependence on modulation format.}
%\color{red}maybe remove "of pulse collisions in a given WDM system".\color{black}
\end{abstract}


\section{Introduction}
The nonlinearity of optical fibers has been long recognized as one of the most important factors limiting the growth of data-rates transmitted in wavelength division multiplexed (WDM) systems \cite{Essiambre}. The nonlinear distortions are usually classified as either intra-channel \cite{intraCh}, or inter-channel \cite{interCh}. While intra-channel distortions are generated by each of the WDM channels individually, inter-channel distortions are caused by interference between different WDM channels. In this work we focus on the physical mechanism responsible for inter-channel nonlinear effects as it comes into play in modern, dispersion uncompensated fiber links with coherent detection. Since joint processing of multiple WDM channels is currently considered to be prohibitively complex in commercial systems, inter-channel interference is customarily treated as noise \cite{Essiambre}, and hence we refer to it in what follows as \emph{nonlinear interference noise}, or NLIN \cite{DarOpex,DarOpex2,DarOL,DarOFC14,DarECOC14,DarJLTInv}.

We follow up on the theory published in \cite{Mecozzi,DarOpex,DarJLTInv} and model in the time domain\footnote{Other time-domain models for inter-channel effects were considered in \cite{Kumar1,Kumar2,Kumar3}} the build-up of NLIN in WDM systems which are using {single-carrier transmission}. In this approach inter-channel NLIN is attributed to multiple pulse collisions whose dynamics we examine in detail (a similar analysis of intra-channel pulse-interactions can be found in \cite{intraCh,MecozziMatera}). As we demonstrate in what follows, the pulse-collision picture provides deep quantitative and qualitative insight into the build-up of NLIN, and explains the various properties of NLIN that have previously only been observed empirically in simulations. For example, issues like the importance of the phase-noise component of the NLIN versus the legitimacy of describing NLIN as complex circular noise, the dependence of the NLIN power on modulation format, the effect of pre-dispersion, the dependence on system-length and type of amplification, etc. \cite{DarOpex,DarOpex2,DarJLTInv,PogECOC,PogEGN,RossiEcoc14,SerenaECOC13}, are all clarified once the essence of pulse collision dynamics is understood.

While the practice of analyzing nonlinear interference by means of pulse collisions has been known for a while (initially in the context of soliton \cite{collisions_soliton,threeSolitons} and dispersion-managed-soliton \cite{collisions_DMS,collisions_DMS2} transmission, and subsequently in the more general context of arbitrary pulse interactions \cite{Shtaif98}), the pulse collision dynamics encountered in modern coherent fiber communications systems, has a number of fundamentally different characteristics. Most prominently, in modern systems, which avoid the use of inline dispersion compensation, simultaneous multi-pulse collisions play a unique and very significant role. In order to illustrate this, consider the case where the channel of interest is perturbed by a nonlinear interaction with a single interfering WDM channel. In the old generation of dispersion managed systems \cite{collisions_DMS,collisions_DMS2,Shtaif98}, interference was generated by collisions between pairs of pulses, one from the channel of interest and one from the interfering channel. In the new generation of dispersion uncompensated systems, pulse spreading generates significant temporal overlap within each of the channels, allowing for example, situations in which two different pulses in the interfering channel interact nonlinearly with a third pulse, belonging to the channel of interest, so as to generate interference that affects a fourth pulse, also belonging to the channel of interest.

%Since the majority of systems operate with standard single-mode fibers that are characterized by fairly high dispersion, the nonlinear distortion resulting from multiple interfering WDM channels is simply the sum of the distortions introduced by the individual interferers, and hence the analysis of a single interferer, which we conduct here, is sufficient in the majority of cases.

%\color{red}Before this paragraph we need to say that we consider only two channels (note that it is mentioned in the next paragraph) and we also need to better explain what we mean by two, three and four pulse-collisions (i.e., two pulse collisions involve a single pulse from the channel of interest and a single pulse from the interfering channel, and so on...).\color{black}

In what follows we demonstrate that substantial understanding of a broad range of observed nonlinear interference phenomena can be achieved by means of proper classification of the type of collisions taking place in the fiber. Most relevantly, there is a qualitative and a quantitative difference between the interference caused by two-pulse, three-pulse and four-pulse interactions. Additionally, the effect of complete collisions (where interacting pulses that belong to different WDM channels pass by each other completely within the fiber span) may differ considerably from the effect of incomplete collisions. In the case of two-pulse collisions, nonlinear interference manifests itself as phase-noise, and polarization-rotation noise, and it is strongly dependent on modulation format. Among the three-pulse collisions some distinctly produce phase and polarization noise, whose variance is modulation format independent, whereas others produce complex circular noise, whose variance is modulation format dependent. Finally, all four-pulse collisions produce complex circular noise whose power is independent of modulation format. While the NLIN produced by two-pulse collisions grows monotonically during the collision process and is maximized when the collision is complete, the behavior of three and four-pulse collisions is distinctly different. The NLIN generated by these collisions is built up constructively in the first half of the collision process and then most of it cancels through destructive interference when the collision is completed. Hence the effect of three and four-pulse collision ends up being much more significant when the collision is incomplete.

The relative importance of complete versus incomplete collisions and therefore the relative importance of two, three and four-pulse collisions, in a given system is determined by the link parameters, predominantly by the length and number of amplified spans, and by the type of amplification that is used. For example, with lumped amplification complete collisions are most significant in short, few-span systems, whereas in the case of distributed amplification (which is the limiting case of Raman amplification \cite{RamanAmp}) the contribution of complete collisions is dominant regardless of the link length. In links dominated by complete collisions, two-pulse collisions are more significant and the NLIN has a distinct phase and polarization noise nature, and it is characterized by a strong dependence on modulation format. As the significance of incomplete collisions increases, four-pulse collisions become more significant so that the NLIN progressively evolves into complex circular noise, and its dependence on modulation format reduces.

Our analysis in what follows focuses only on nonlinear interference between two WDM channels, one of which is referred to as the channel of interest, and the other is referred to as the interfering channel. In the traditional jargon of WDM systems \cite{ForgBookCh}, this would qualify as studying only the effect of cross-phase-modulation (XPM), and omitting inter-channel four wave mixing (FWM) processes in which three or four different WDM channels are involved. As is evident from simulations \cite{DarOpex,PogEGN}, the contribution of FWM to NLIN in most relevant fiber types is of negligible importance\footnote{{As pointed out in} \cite{PogEGN}, {in certain types of non-zero-dispersion-shifted fibers, one can observe a contribution to NLIN that originates from FWM processes. However, even in these situations, the FWM contribution is a relatively small correction to the contribution of XPM, as can be further verified in} \url{http://nlinwizard.eng.tau.ac.il}.}. Under these circumstances, the NLIN that is generated by multiple WDM channels is simply the sum of the NLIN contributions produced by the individual interferers. {Furthermore, in this work we concentrate on a regime in which the nonlinear interaction between the signal and the amplified spontaneous emission (ASE) noise is negligible.}

%Finally, we stress that in our definition, a complete collision between pulses occurs when the evolution of the individual pulse waveforms during the collision is negligible. This condition can be translated into two requirements. The first is that the dispersive broadening of the pulses during the collision is negligible relative to the pulse width prior to the collisions. The second is that the effect of attenuation is negligible while the collision is taking place. It can be shown that the first requirement is automatically satisfied when the frequency spacing between the interacting WDM channels is sufficiently larger than the bandwidth of the channels themselves. The second requirement is more limiting in the setting of a modern dispersion uncompensated link. The reason is that owing to the link's accumulated dispersion, the pulse duration increases during propagation, and hence the length of the fiber section in which an individual collision takes place is eventually bound to reach values over which the effect of fiber attenuation cannot be neglected. Therefore, true complete collisions exist in relatively short links (where the pulses remain sufficiently short), or in those using distributed amplification. Nevertheless, as we will demonstrate in what follows, the existence of complete collisions explains all of the observed conundrums related to nonlinear noise accumulation.

The paper is organized as follows. In Sec. \ref{TDG} we briefly review the essentials of the time domain model, initially introduced in \cite{Mecozzi,DarOpex,DarJLTInv} and establish the background for the subsequent analysis. In Secs. \ref{TwoPulse}, \ref{ThreePulse}, and \ref{FourPulse} we classify the nonlinear interference processes as two, three and four-pulse collisions, respectively. In each case we describe their distinctive features depending on whether the collision is complete, or incomplete. Several numerical examples illustrating the main ideas are presented in Sec. \ref{numerics}. For the simplicity of illustration Secs. \ref{TDG}--\ref{FourPulse} consider scalar (single-polarization) transmission. The extension to the polarization multiplexed case is presented in Sec. \ref{polarization}. %Polarization multiplexed transmission is also assumed in all the sections that follow.
In Sec. \ref{System} we discuss the  implications of the pulse-collision theory to fiber-communications systems, whereas in Sec. \ref{GN} we explain the role of chromatic dispersion and relate the time domain theory to the Gaussian noise (GN) model \cite{Poggiolini,Carena,Pog2,Johannisson,Bononi}. Section \ref{conc} is devoted to conclusions.

\begin{figure*}[t]
\center
\includegraphics[width=1.0\textwidth]{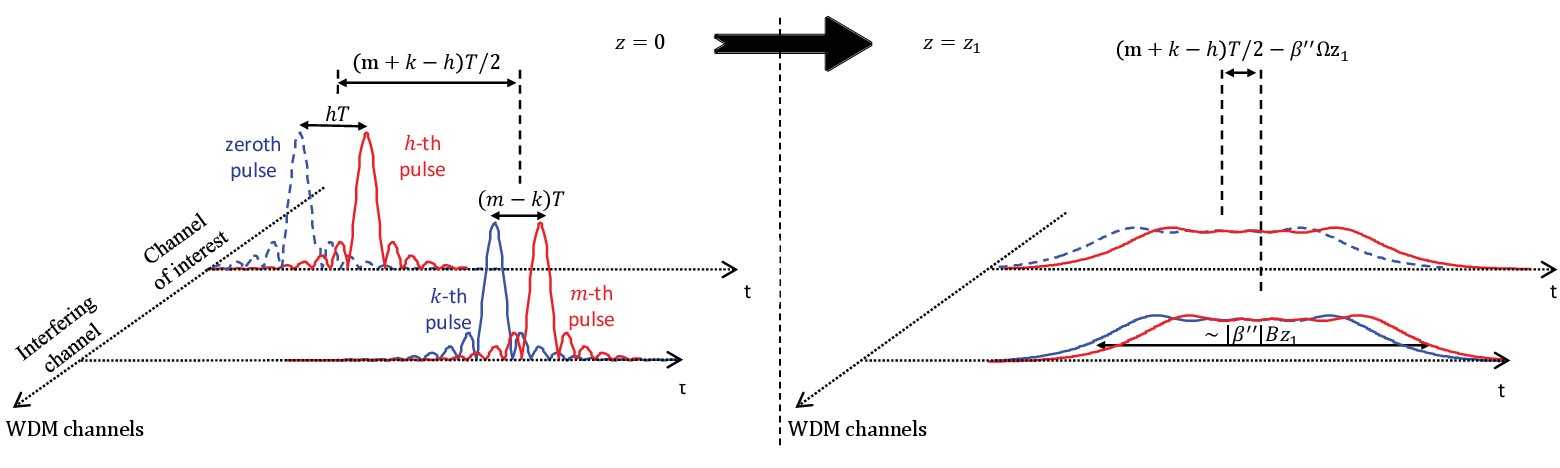}
%\vspace{-0.4cm}
\caption{The generic four-pulse interference. The $h$-th pulse in the channel of interest interacts with the $k$-th and $m$-th pulses of the interfering channel so as to create NLIN that affects the zeroth symbol in the channel of interest. Since we assume detection that involves filtering matched to the waveform of the fundamental pulse centered at the zeroth symbol, as discussed in Sec. \ref{TDG}, one can refer to this as a four-pulse interaction. The zeroth pulse is plotted only to keep track of the matched-filter waveform, and it is plotted by a dashed line in order to stress that it is not a pulse that is physically participating in the nonlinear interaction during propagation. The left panel shows the situation in the beginning of the link, before the pulses overlap, whereas the right panel illustrates the situation once an overlap between the pulses is formed. When two of the indices overlap ($h=0$, or $k=m$) one observes a three-pulse interaction and when both index pairs overlap ($h=0$ and $k=m$) one obtains a two-pulse interaction.}\label{Figure1}
\end{figure*}

\section{Time-domain theory\label{TDG}}
For the simplicity of illustration we start by considering the single-polarization case, postponing the generalization to polarization multiplexed systems to Sec. \ref{polarization}.
In the case of two WDM channels, we may express the electric field at the fiber input as
\bea \sum_n a_ng(0,t-nT) + \exp(-i\Omega t)\sum_n b_n g(0,t-nT),\eea
where $T$ is the symbol duration and $a_n$ and $b_n$ are the complex valued data symbols transmitted over the channel of interest and over the interfering channel, respectively. The central frequency of the channel of interest is arbitrarily set to 0, whereas the central frequency of the interfering channel is denoted by $\Omega$. The injected fundamental symbol waveforms in both channels are identical and denoted by $g(0,t)$, whereas $g(z,t)=\exp(-\frac{i}{2}{\beta''}z\frac{\partial^2}{\partial t^2})g(0,t)$ is the dispersed waveform of the individual pulse when reaching point $z$ along the fiber with $\beta''$ being the fiber dispersion coefficient\footnote{As we are relying on a first order perturbation analysis -- the waveform $g(z,t)$ includes only the effect of dispersion and not the effect of nonlinearity. In \cite{DarOpex} we used $g^{(0)}(z,t)$ for the same quantity.}. The fundamental waveforms are further assumed to be orthogonal with respect to time shifts by an integer number of symbol durations and their energy is assumed to be normalized to 1
\bea \int g^*(z,t-nT)g(z,t-n'T) \df t=\delta_{n,n'}\label{eqA10}.\eea
After coherent detection, the channel of interest is match-filtered with the filter's impulse response being proportional to $g^*(L,t)$, where $L$ is the link length. The extracted $n$-th data symbol can  be then expressed as $a_n+\Delta a_n$, with $\Delta a_n$ accounting for the presence of NLIN. In \cite{DarOpex} we have shown that $\Delta a_n$ due to inter-channel NLIN  is given by
\bea \Delta a_0 \hspace{-0.3cm}&=&\hspace{-0.3cm} 2i \gamma \sum_{h,k,m}  a_h b_k^* b_m   X_{h,k,m}. \label{eq60} \eea
where $\gamma$ is the nonlinear coefficient and where we have arbitrarily (and without loss of generality) set the index of the received symbol to $0$. The coefficients $X_{h,k,m}$ have been introduced in Ref. \cite{DarOpex}, where they have been shown to be
\bea X_{h,k,m}\hspace{-0.3cm}&=&\hspace{-0.3cm} \int_0^L\hspace{-0.3cm} \df z ~f(z)\int_{-\infty}^\infty\hspace{-0.1cm} \df t ~g^{*}(z,t) g(z,t-hT)\nonumber\\
 &&\hspace{-0.3cm} \times g^{*}\hspace{-0.0cm}(z,t\hspace{-0.1cm}-\hspace{-0.1cm}kT\hspace{-0.1cm}-\hspace{-0.1cm} \beta'' \Omega z) g(z,t\hspace{-0.1cm}-\hspace{-0.1cm}mT\hspace{-0.1cm}-\hspace{-0.1cm} \beta'' \Omega z).\label{eq80}\eea
The function $f(z)$ accounts for the loss/gain profile along the link \cite{Mecozzi,DarOpex}. For example, $f(z)=1$ in the case of perfectly distributed amplification and $f(z)=\exp(-\alpha\hspace{-0.2cm}\mod\hspace{-0.1cm}(z,L_s))$ in the case of lumped amplification where $\alpha$ is the loss coefficient and $L_s$ is the span length.

The coefficient $X_{h,k,m}$ is associated with the NLIN that is observed in the measurement of the zeroth data symbol in the channel of interest following the nonlinear interaction between the $h$-th pulse in the channel of interest and the $k$-th and $m$-th pulses in the interfering channel, as illustrated in Fig. \ref{Figure1}. The triple product $\xi(z,t) = g(z,t-hT)g^{*}\hspace{-0.1cm}(z,t\hspace{-0.1cm}-\hspace{-0.1cm}kT\hspace{-0.1cm}-\hspace{-0.1cm} \beta'' \Omega z) g(z,t\hspace{-0.1cm}-\hspace{-0.1cm}mT\hspace{-0.1cm}-\hspace{-0.1cm} \beta'' \Omega z)$, appearing in the integrand of Eq. \eqref{eq80}, describes a classic nonlinear Kerr interaction taking place at time $t$ and position $z$ along the link (where $\beta'' \Omega$ is the difference between the reciprocal group velocities). This product then propagates towards the receiver, while accumulating the dispersion present in the remaining fiber-length, which is equal to $L-z$. Since the impulse response of the matched filter is proportional to $g^*(L,t)=\exp(\frac{i}{2}{\beta''}(L-z)\frac{\partial^2}{\partial t^2})g^*(z,t)$, this accumulated dispersion is removed. The sampled perturbation at the receiver, as the inner integral in Eq. \eqref{eq80} suggests, is therefore $\int g^*(z,t) \xi(z,t)\df t$, which can be interpreted as if the nonlinear product $\xi(z,t)$ was matched filtered by a `fourth' pulse $g^*(z,t)$ exactly at the position along the link at which the Kerr interaction occurred. The integral over $z$ in Eq. \eqref{eq80} gathers the nonlinear products produced along the entire link.

We now proceed to classifying the various contributions to the NLIN in terms of the type of collisions that produce them. In the most general case,  these contributions are produced by nonlinear interactions between four pulses -- the ``dummy'' zeroth pulse from the channel of interest (accounting for the fact that the signal is matched filtered at the receiver), the $h$-th pulse from the channel of interest, and the $k$-th and $m$-th pulses from the interfering channel. When some of the indices coincide, as we shall see in the following sections, a two-pulse, or three-pulse collision is formed.

\section{Two-pulse collisions\label{TwoPulse}}
A two-pulse collision occurs when one of the interacting pulses is the pulse of interest, which we arbitrarily select as the zeroth-index pulse, whereas the other pulse belongs to the interfering channel, so that $h=0$ and $k=m$. The strength of this interaction is governed by the coefficient
\bea X_{0,m,m}\hspace{-0.3cm}&=&\hspace{-0.3cm} \int_0^L\hspace{-0.3cm} \df z f(z)\hspace{-0.1cm}\int_{-\infty}^\infty\hspace{-0.4cm} \df t \abs{g(z,t)}^2 \abs{g(z,t\hspace{-0.1cm}-\hspace{-0.1cm}mT\hspace{-0.1cm}-\hspace{-0.1cm} \beta'' \Omega z)}^2\hspace{-0.1cm},\label{eq89}\eea
and, according to Eq. \eqref{eq60}, it produces a perturbation equal to $i a_0 (2 \gamma X_{0,m,m}|b_m|^2)$. As $X_{0,m,m}$ is a real-valued quantity the perturbation is at a complex angle of $\pi/2$ from the transmitted symbol $a_0$. Namely, the perturbation generated by two-pulse interactions has the character of phase-noise.

An interesting perspective into this reality can be obtained by considering the `old class' of two-pulse collisions initially studied in the context of solitons \cite{collisions_soliton,collisions_DMS,collisions_DMS2,threeSolitons}. There, it has become common knowledge that when a pulse in the channel of interest undergoes a complete collision with an interfering pulse belonging to an adjacent WDM channel, the only notable consequence observed on the pulse of interest is a \emph{time-independent} phase-shift. However, it is also well known that when the collision between the two pulses is not complete \cite{Shtaif98,collisions_DMS2}, the phase of the interacting pulses is distorted via XPM leading to the formation of a \emph{time-dependent} phase distortion (nonlinear chirp), which subsequently produces an amplitude (or timing-jitter) perturbation after being affected by the dispersion of the remaining fiber. Note that in the scheme that we consider here, no amplitude perturbation is formed in two-pulse collisions, regardless of whether the collisions are complete, or incomplete. This difference with respect to the old generation of systems can be attributed to coherent detection and the use of a matched filter, which is equivalent to measuring the perturbation at the point along the fiber where the collision takes place, as explained earlier. Namely, the effect of dispersion, which translates the time-dependent phase distortion into amplitude variations is reversed by the matched filter.

Since in addition to being real-valued, the integrand in Eq. \eqref{eq89} is also non-negative, the nonlinear phase-shift is proportional to the ``completeness'' of the collision, namely the phase-shift is zero when the waveforms do not overlap at all during propagation and it is maximized when the collision is complete.
%This behavior can be observed in Fig. \ref{Figure2}a and Fig. \ref{Figure2}b, showing (red curves) the real and imaginary parts of the coefficient $X_{0,?,?}$, respectively. Also shown in the figure, in addition to the coefficients themselves, are the curves representing the inner integral in Eq. \eqref{eq80} which represents the contribution of a single fiber increment $\df z$ to the overall NLIN. As we pointed out earlier, it is the positivity of the imaginary part of the incremental contributions in the case of two-pulse collisions that ensures the monotonic increase of the overall two-pulse NLIN with $z$.
As we demonstrate in Appendix \ref{Collisions}, two-pulse collision coefficients $X_{0,m,m}$ scale as $\Omega^{-1}$. In addition, the fact that two-pulse contributions to NLIN are proportional to $a_0|b_m|^2$ implies that the variance of NLIN that is associated with these contributions is proportional to $\lip |a_0|^2\rip(\lip|b_m|^4\rip-\lip |b_m|^2\rip^2)$, with the angled brackets denoting statistical averaging. While $\lip |a_0|^2\rip$ is simply proportional to the average signal power $\lip|b_m|^4\rip-\lip |b_m|^2\rip^2$, is clearly dependent on modulation format. For example, when the interfering channel is modulated only in phase (e.g. QPSK) the NLIN variance caused by two-pulse collisions is 0.

\begin{figure*}[t]
\center
\includegraphics[width=1.0\textwidth]{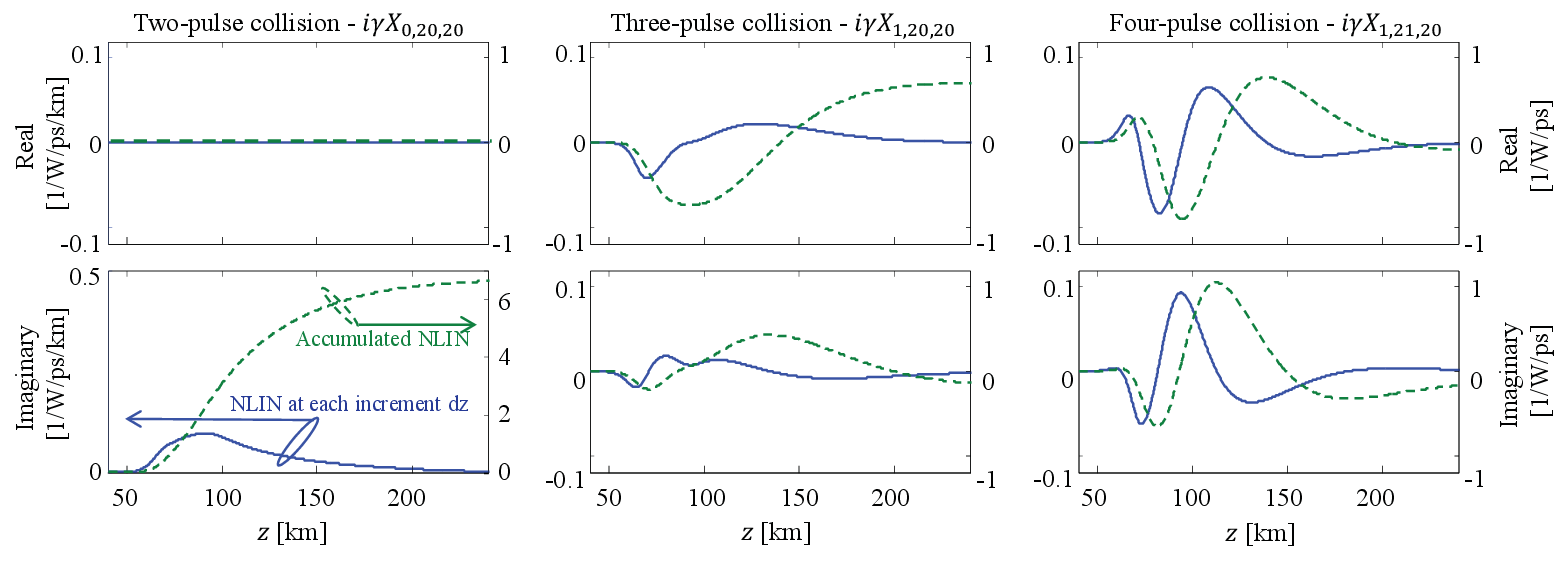}
%\vspace{-0.4cm}
\caption{The evolution of the NLIN coefficients along the fiber axis $z$, assuming {ideally} distributed amplification {and channel separation of $50$GHz}. The solid (blue) curves show the contribution from a single fiber increment $\df z$ at position $z$ along the fiber to the NLIN component $i\gamma X_{h,k,m}$ (this is proportional to the inner integral in Eq. (\ref{eq80})), whereas the dashed (green) curves show the accumulated NLIN component $i\gamma X_{h,k,m}$ itself, under the assumption that the link is ended at point $z$. The two-pulse component $i\gamma X_{0,20,20}$ is imaginary and hence produces pure phase-noise. It grows monotonically with distance, reaching a maximum after the collision has been completed. The three and four-pulse components $i\gamma X_{1,20,20}$ and $i\gamma X_{1,21,20}$ are complex and their value is largest during the collision itself and reduces significantly when the collision is complete.} \label{Figure2}
\end{figure*}

\section{Three-pulse collisions\label{ThreePulse}}
Three-pulse interactions are represented by coefficients of the form $X_{0,k,m}$ (where $k\neq m$) and $X_{h,m,m}$ (where $h\neq 0$). The first form represents an interaction between the pulse of interest (the zeroth pulse from the channel of interest) and the $k$-th and $m$-th pulses from the interfering channel. The perturbation produced by this interaction is equal to $i a_0(2\gamma X_{0,k,m}b_k^* b_m + 2\gamma X_{0,m,k}b_m^* b_k)$ and since it can be easily verified that $X_{0,k,m}=X^*_{0,m,k}$, the perturbation is equal to $i a_0 \mathfrak{Re} \{ 4 \gamma X_{0,k,m}b_k^* b_m \}$ which is at a complex angle of $\pi/2$ from the transmitted symbol $a_0$ and therefore it too has the character of phase-noise. Yet, unlike the case of two-pulse collisions, since $k\neq m$ the variance of this phase-noise contribution depends only on the average signal power and not on the modulation format. The second form of three-pulse interactions $X_{h,m,m}$ involves an interaction between two pulses in the channel of interest (the zeroth and $h$-th pulses) and a single pulse from the interfering channel (the $m$-th pulse). The perturbation produced by this interaction is equal to $\Delta a_0 =i a_h (2 \gamma X_{h,m,m}|b_m|^2)$. This contribution has no fixed phase relation with $a_0$ and hence it can be viewed as complex circular noise. Yet, similarly to the case of two-pulse collisions, the dependence on $|b_m|^2$ implies dependence of the NLIN variance due to this contribution on the data modulation format.

We show in Sec. \ref{numerics} and in Appendix \ref{Collisions} that the perturbation to the symbol of interest is formed within the initial part of a three-pulse collision, but then cancels through destructive  interference once the collision is complete, so that the residual magnitude of the perturbation scales as $\Omega^{-2}$. In the regime where the relative number of complete collisions is significant, the overall contribution of three-pulse collisions practically vanishes relative to that of two-pulse collisions (in spite of the fact that the number of three-pulse collisions is larger, see Sec. \ref{System}). The scaling of incomplete three-pulse collisions with frequency is not monotonous, and depends on the point at which the collision is discontinued.

\section{Four-pulse collisions\label{FourPulse}}
The case of four-pulse collisions involves two pulses from the channel of interest (the zeroth and $h$-th pulses, with $h\neq 0$) and two pulses from the interfering channel (the $k$-th and $m$-th pulses with $k\neq m$). Each of these interactions generates a perturbation that is equal to $2i\gamma a_h ( X_{h,k,m}b_k^* b_m +  X_{h,m,k}b_m^* b_k)$ and since there is no fixed phase relation between the various symbols, these contribution can be viewed as producing complex circular noise. Since $h\neq 0$ and $k\neq m$, and the data transmitted on different symbols is assumed to be statistically independent, the average value of NLIN due to four-pulse contributions is 0, and the variance is proportional to $\lip |a_0|^2\rip \lip |b_0|^2\rip^2$, which is determined by the average power per symbol independently of the modulation format.

We show in Sec. \ref{numerics} and in Appendix \ref{Collisions} that as in the case of three-pulse collisions, the interference formed in the first half of a four-pulse collision process cancels through destructive interference in the second half. The magnitude of the interference produced by a four-pulse collision is therefore determined by the point in which the collision is discontinued. In Appendix \ref{Collisions} we show that the maximal interference produced by an incomplete collision scales as $\Omega^{-1}$ whereas the magnitude of the residual interference left after a complete collision scales as $\Omega^{-3}$. This implies that the overall contribution of complete four-pulse collisions is negligible in spite of their much larger number (see Sec. \ref{System}). %With respect to incomplete collisions, four-pulse collisions produce a significant contribution because their number is much larger than that of incomplete two, or three pulse collisions, and because all types of incomplete collisions scale similarly with the frequency separation $\Omega$.

\section{A few numerical examples\label{numerics}}
In order to better illustrate the distinctive features of the various types of collisions we first examine the dependence of three particular NLIN coefficients on space and frequency separation. These coefficients are $X_{0,20,20}$, $X_{1,20,20}$ and $X_{1,21,20}$, representing a two-pulse, three-pulse and four-pulse collision, respectively. {Notice that $X_{1,20,20}$ can be viewed as representing both types of three-pulse collisions (i.e., those with $h=0$ and those with $h\neq 0$), as it can be easily shown that $X_{h,m,m}=X_{0,m-h,m}^*$.} Since the concept of complete collisions is best isolated when losses during the collision are negligible, we plot the examples in this section in the {ideally} distributed amplification case. The issue of lumped versus distributed amplification is further discussed in Sec. \ref{System}. We assume square-root raised-cosine pulses of $32$GHz baud-rate with a roll-off factor of 0.2, {transmitted over a standard single-mode fiber with dispersion coefficient $\beta''=21$ps$^2$/km, nonlinear coefficient $\gamma=1.3$ W$^{-1}$km$^{-1}$ and attenuation of 0.2 dB/km}.

In Fig. \ref{Figure2} we {assume channel separation of $50$GHz and} show the evolution of the various NLIN coefficients along the fiber, where the real and imaginary parts of the perturbation $i\gamma X_{h,k,m}$ are plotted in the top and bottom panels, respectively. In each figure the solid (blue) curve shows the contribution to the NLIN produced in a single fiber increment of length $\df z$ (the inner (time) integral of Eq. \eqref{eq80}). The dashed (green) curves show the accumulated contribution to the NLIN, obtained by integrating the corresponding solid (blue) curves from the link input up to point $z$ along the fiber. In the case of two-pulse collisions, as we have argued earlier, the incremental contribution is imaginary (i.e. it is in quadrature with the symbol of interest), and hence it only contributes to the formation of phase-noise. In addition, the imaginary part of the incremental contributions is strictly non-negative and hence, the corresponding phase-perturbation increases monotonically with position along the link. Therefore, in the case of an incomplete collision (where the integration ends while the collision is still taking place) the phase-noise is smaller than it is when the collision is complete. With three and four-pulse collisions the situation is distinctly different. First, the incremental contributions are complex, and secondly, the incremental contributions change signs during the collision, so that the integrated NLIN coefficient grows in the beginning of the collision, but then reduces towards its end. The fact that with three and four-pulse collisions, the end-value of the accumulated contribution (green-dashed curves) is lower than the corresponding peak value (around the center of the collision) is clearly evident in the center and right panels of Fig. \ref{Figure2}, suggesting that the interference induced by incomplete collisions is likely to be much stronger than it is in the case of complete collisions. {This feature is particularly strong in the case of four-pulse collisions and for both three-pulse and four-pulse collisions it becomes increasingly more distinct when the frequency separation between the interacting WDM channels becomes larger.}

\begin{figure}[t]
\center
\includegraphics[width=0.45\textwidth]{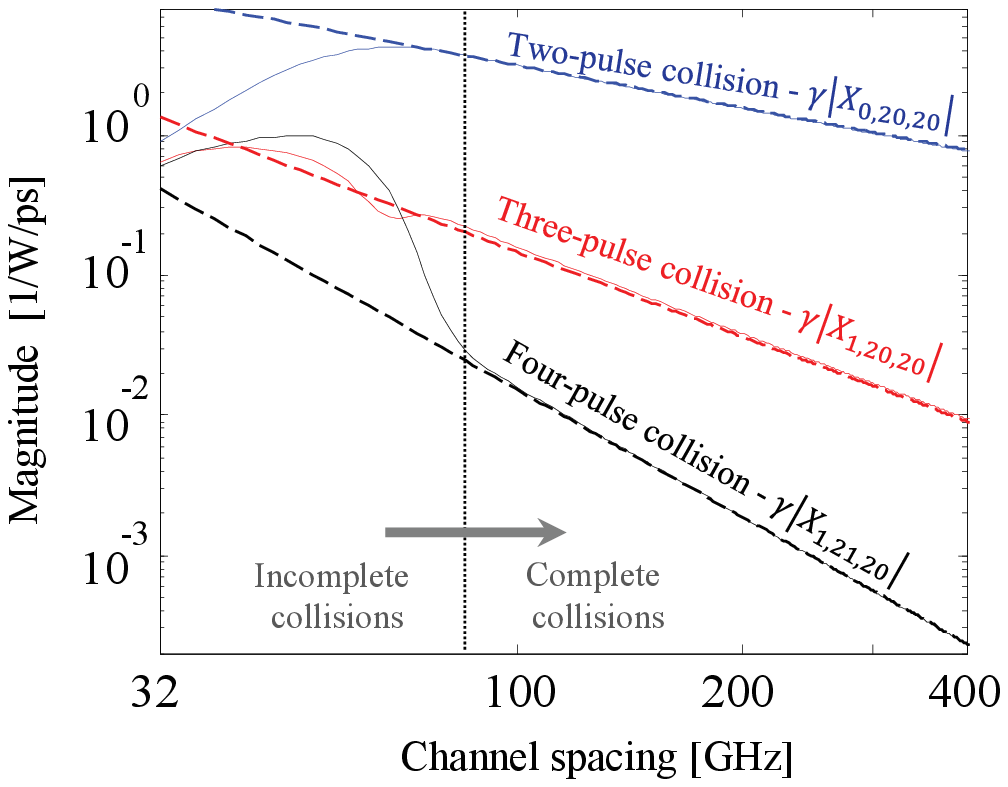}
%\vspace{-0.4cm}
\caption{The magnitude of the nonlinear perturbation $\gamma|X_{0,20,20}|$, $\gamma|X_{1,20,20}|$, and $\gamma|X_{1,21,20}|$ as a function of frequency separation for a $100$ km link with {ideally} distributed amplification. It can be clearly observed that once the collision becomes complete (according to Eq. \eqref{chSp}, this happens when the channel separation is larger than $85$GHz), the two-pulse NLIN coefficient drops as $\Omega^{-1}$, the three-pulse coefficient drops as $\Omega^{-2}$ and the four-pulse coefficient drops as $\Omega^{-3}$ (dashed lines show the scaling with $\Omega^{-j}$ for $j=1,2$ and $3$). Note that for longer links, the transition from incomplete to complete collisions would take place at a correspondingly lower channel spacing.}\label{Figure3}
\end{figure}

%The tails level, i.e. the value of the NLIN coefficient after the collision completes, reduces as $\Omega^{-1}$ in the case of two-pulse collisions, as $\Omega^{-2}$ in the case of three pulse collisions and as $\Omega^{-3}$ in the case of four-pulse collisions, consistent with the fact that the tails level in the rightmost panels of Fig. \ref{Figure2} is indeed the lowest.
The dependence on frequency separation is illustrated in Fig. \ref{Figure3} where we show the dependence of the coefficients $X_{h,k,m}$ on $\Omega$ for a 100 km link. As can be deduced from Appendix \ref{Collisions}, when
\bea \Omega>B+\frac{(m+k-h)T}{2|\beta''|L}, \label{chSp}\eea
{where $B$ is the spectral width}\footnote{We stress that in Eq. \eqref{chSp} and throughout this paper $\Omega$ and $B$ are indicated in angular frequency units.} {of the individual pulses,} the group velocity differences between the channels are large enough to guarantee that the collision is completed before the fiber's end. In the case considered in Fig. \ref{Figure3}, this happens when the channel spacing is larger than $85$GHz. Evidently, in this regime the perturbation due to two, three, and four-pulse collisions scales as $\Omega^{-1}$, $\Omega^{-2}$, and $\Omega^{-3}$, respectively. When the frequency separation is below $85$GHz, the collision is no longer complete, as it is discontinued abruptly at the end of the fiber. In this regime the magnitude of the perturbation depends on the point in which the collision is disrupted and hence, consistently with the discussion of Fig. \ref{Figure2}, the dependence on $\Omega$ is only monotonous in the two-pulse collision case.

\begin{figure}[t]
\center
\includegraphics[width=0.45\textwidth]{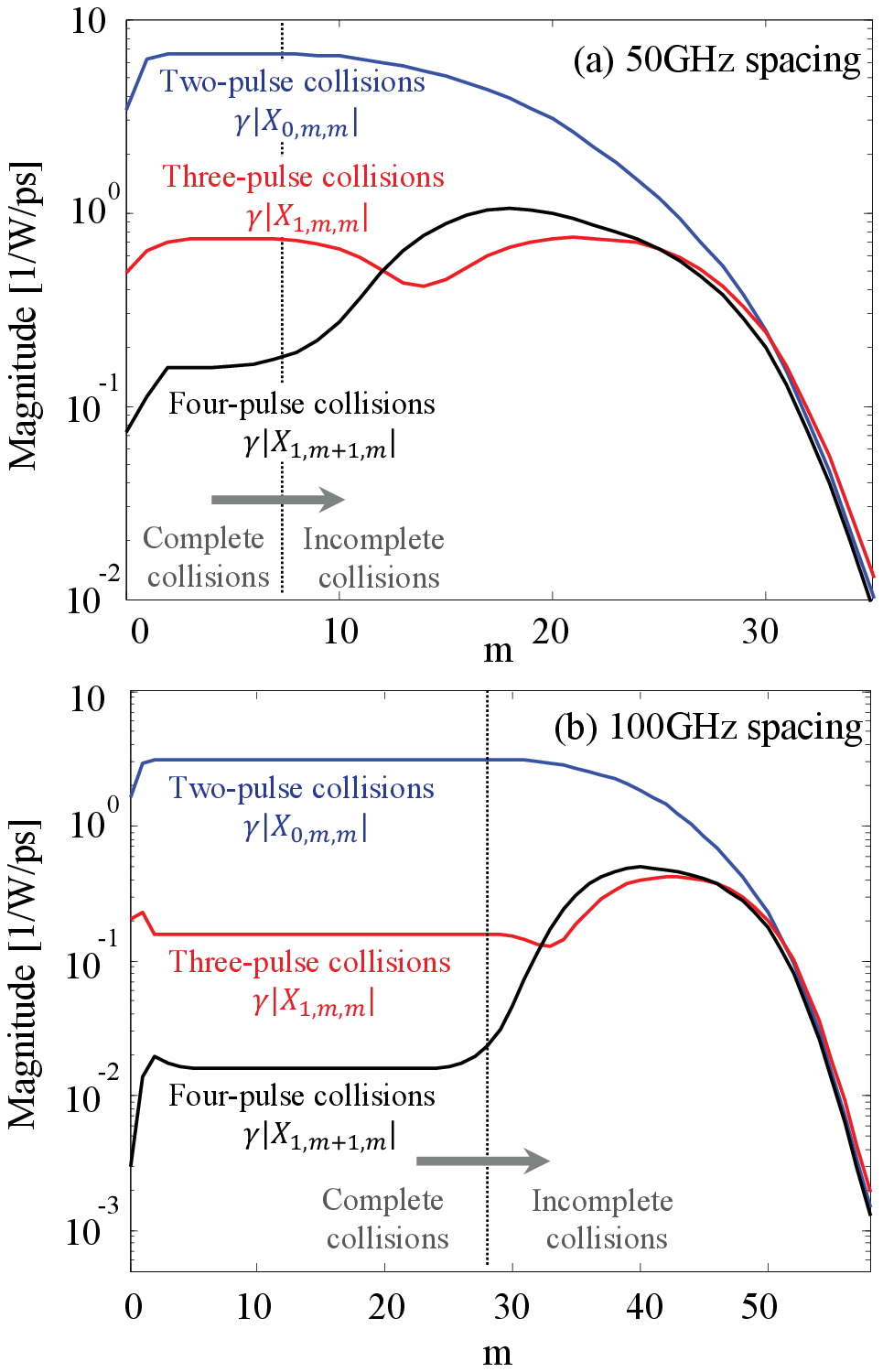}
%\vspace{-0.4cm}
\caption{{The magnitude of the nonlinear perturbation $\gamma|X_{0,m,m}|$, $\gamma|X_{1,m,m}|$, and $\gamma|X_{1,m+1,m}|$ as a function of $m$ for a $100$ km link with ideally distributed amplification. Top and bottom panels assume $50$GHz and $100$GHz channel separations, accordingly. In longer links, the transition from incomplete to complete collisions would take place at a correspondingly higher values of $m$.%When the collisions are complete (as can be deduced from Appendix \ref{Collisions}, this happens when $m\leq 7$ in the case of $\Omega=50z$GH and $m\leq 28$ in the case of $\Omega=100$GHz) two-pulse collisions receive their maximal value, whereas three-pulse and four-pulse collisions induce an interference that is one and two orders of magnitude below that value, respectively.
}}\label{scalingWithM}
\end{figure}

{We further illustrate the properties of the various types of collisions by examining their dependence on the index of the interfering pulses while fixing the channel separation and the link length. In Fig. \ref{scalingWithM} we plot the magnitude of $i\gamma X_{0,m,m}$, $i\gamma X_{1,m,m}$ and $i\gamma X_{1,m+1,m}$ versus $m$ in a $100km$ link. Figure \ref{scalingWithM}a shows the case of $50$GHz channel spacing whereas Fig. \ref{scalingWithM}b shows the case of $100$GHz channel spacing. As can be deduced from Eq. \eqref{chSp}, the collisions are complete when $m< 7$ for channel separation of $50$GHz and $m<28$ for channel separation of $100$GHz. In this regime $\gamma|X_{0,m,m}|$ receives its maximum value, while $\gamma|X_{1,m,m}|$ and $\gamma|X_{1,m+1,m}|$ are approximately one and two orders of magnitude below that value, respectively. When $m$ increases, the collisions become incomplete and the perturbation in the case of two-pulse collisions reduces monotonically while it oscillates in the cases of three-pulse and four-pulse collisions. Eventually, when $m$ grows further the pulses do not collide at all and all of the coefficients vanish}\footnote{{We note that the reasoning presented in this paper does not strictly apply to the cases of very low $m$ values (such as $m=0$, or $m=1$ in Fig. \ref{scalingWithM}), where more than two-pulse collisions cannot be formed for lack of sufficient dispersion.}}.

%As can be deduced from the appendix, full collisions occur when $\Omega>85$ GHz, where the magnitude of the NLIN coefficients $X_{0,20,20}$, $X_{1,20,20}$, and $X_{1,21,20}$ reduces as $\Omega^{-1}$, $\Omega^{-2}$, and $\Omega^{-3}$, respectively.  This is consistent with the fact that these coefficients corresponds to two, three and four-pulse interactions, as explained earlier. With smaller values of $\Omega$ the collision take place towards the link's end and are hence incomplete. In the case of incomplete collisions the dependence on $\Omega$ becomes complicated, as the reduction of $\Omega$ extends the collision length and shifts the center of the collision closer to the fiber end. In fact, when $\Omega$ is reduced further all three coefficients will eventually drop to zero when the end of the link arrives sooner than the overlap between the pulses is formed.

%

\section{The effect of polarization-multiplexing}\label{polarization}

\begin{table*}[t]
\center
\includegraphics[height=0.25\textwidth,width=0.9\textwidth]{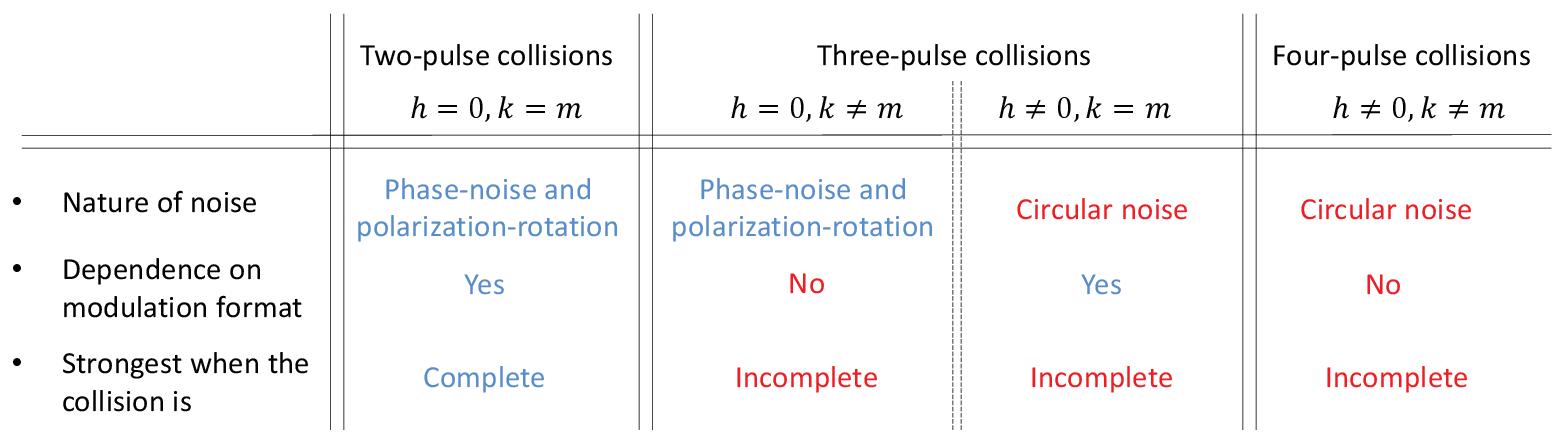}
\caption{Summary of XPM contributions}
\label{Table}
\end{table*}

Throughout Secs. \ref{TwoPulse}-\ref{FourPulse} we have assumed for simplicity the case of a signal transmitted in a single polarization. While the consideration of polarization multiplexed transmission is considerably more cumbersome, the principles remain very similar. Equation (\ref{eq60}) is generalized to
\bea \Delta \uu a_0 \hspace{-0.3cm}&=&\hspace{-0.3cm} i \gamma \sum_{h,k,m}   X_{h,k,m} \left(\uu b_k^\dagger \uu b_m \mathbf I+ \uu b_m\uu b_k^\dagger\right)   \uu a_h, \label{eq60P} \eea
where the underline denotes a two-element column vector (i.e. its elements represent the two polarization component of the optical field) and $\mathbf I$ is the $2\times 2$ identity matrix. For simplicity Eq. (\ref{eq60P}) corresponds to the case in which the modulated polarization axes in the interfering channel are parallel to those in the channel of interest. The more general case, in which the modulated polarization axes in the two WDM channels are rotated relative to each other is treated in Appendix \ref{PolRot}. The NLIN in Eq. (\ref{eq60P}) is identical to that given in \eqref{eq60}, except that each pulse is now modulated by a vector and carries two complex degrees of freedom. The coefficient $X_{h,k,m}$ given in Eq. (\ref{eq80}) remains unchanged, and hence its scaling with the frequency separation $\Omega$ is also unaltered.

As in the case of single-polarization, interference terms that are proportional to coefficients of the form $X_{0,k,m}$ (corresponding to either two-pulse collisions (when $k=m$) or three-pulse collisions (when $k\neq m$)) do not change the norm of the detected vector of symbols $\uu a_0$, but they do change its phase and state of polarization. Thus, the phase-noise content of NLIN in the single-polarization case is generalized into phase-noise \textit{and} polarization-rotation. We first show this for the case of two-pulse collisions, whose contribution to the NLIN is given by
\bea \Delta \uu a^{(\mathrm{2PC})}_{0} \hspace{-0.3cm}&=&\hspace{-0.3cm} i \gamma \sum_{m}   X_{0,m,m} \left(\|\uu b_m\|^2 \mathbf I+ \uu b_m\uu b_m^\dagger\right)   \uu a_0 \nonumber\\
\hspace{-0.3cm}&=&\hspace{-0.3cm} \frac{i}{2} \gamma \sum_{m}   X_{0,m,m} \left(3\|\uu b_m\|^2 \mathbf I+ \vec{B}_m\cdot\vec{\sigma}\right)   \uu a_0,
\label{eq60PP} \eea
where the superscript `2PC' stands for two-pulse-collisions, and where $\|\uu b_m\|^2 = \uu b_m^\dagger\uu b_m$ is the square modulus of the $m$-th vector of symbols. The Second line in Eq. (\ref{eq60PP}) uses the Stokes-space representation \cite{GordonPNAS}. The term $\vec\sigma=(\sigma_1,\sigma_2,\sigma_3)$ is a vector whose three components are the Pauli matrices. The term $\vec B_m$ is the vector that represents $\uu b_m$ in the Stokes space. Its three components are real-valued and given by  $B_{m,j} = \uu b_m^\dagger\sigma_j\uu b_m$, with $j=1,2,$ or $3$. The notation $\vec B_m\cdot\vec\sigma$ is a short-hand for $B_{m,1}\sigma_1+B_{m,2}\sigma_2+B_{m,3}\sigma_3$.  All of the relevant properties of the Stokes representation are conveniently summarized in  \cite{GordonPNAS}. Within the first order perturbation analysis, the perturbed vector can be written as\footnote{Since the entire theory relies on the first-order perturbation analysis, Eq. (\ref{pRot}) is not an approximation of Eq. (\ref{eq60PP}), but rather it is just as accurate as Eq. (\ref{eq60PP}) is.}
\be \uu a_0 + \Delta \uu a^{(\mathrm{2PC})}_{0} = e^{i\varphi_{\mathrm{2PC}}} e^{\frac{i}{2}\vec S_{\mathrm{2PC}}\cdot\vec{\sigma}}\uu a_0.\label{pRot}
\ee
where
\bea \varphi_{\mathrm{2PC}} &=& \frac{3}{2}\gamma \sum_{m}   X_{0,m,m} \|\uu b_m\|^2\label{phi2}\\
\vec S_{\mathrm{2PC}} &=& \gamma \sum_m X_{0,m,m}\vec{B}_m.\label{S2}\eea
While the first exponent in Eq. (\ref{pRot}) represents a phase rotation of both polarizations by an angle $\varphi_{\mathrm{2PC}}$, the second exponent represents a polarization-rotation. In the Stokes jargon, the vector representing $\uu a_0$ on the Poincar\'{e} sphere, rotates about the vector $\vec S_{\mathrm{2PC}}$ at an angle equal to its length $|\vec S_{\mathrm{2PC}}|$ \cite{GordonPNAS}.

As can be readily deduced from Eq. \eqref{phi2}, the phase-noise $\varphi_{\mathrm{2PC}}$ strongly depends on the modulation format, and in particular, its variance vanishes in the case of phase-modulated transmission (such as QPSK) as $\|\uu b_m\|^2$ is constant. The dependence of the polarization rotating effect on modulation format can be seen by observing how the orientation of the vector $\vec{B}_m$ is scattered on the Poincar\'{e} sphere. In particular, the isotropy of the orientations of $\vec B_m$ on the Poincar\'{e} sphere increases with increasing QAM-order, whereas with BPSK, or QPSK, all the vectors $\vec{B}_m$ reside in a single plane. Since the polarization-rotation is determined by $\sum_m X_{0,m,m}\vec{B}_m$, the variance of the rotations angle $|\vec S_{\mathrm{2PC}}|$ also depends on modulation format (as shown in the section that follows), but it is always non-zero, even in the case of pure phase-modulation.

The effect of three-pulse collisions of the type $h=0$ and $k\neq m$ generalizes identically to the above, except that $\varphi_{\mathrm{2PC}}$ and $\vec S_{\mathrm{2PC}}$ are replaced by
\bea \varphi_{\mathrm{3PC}} &=& \frac{3}{2}\gamma \sum_{\substack{k,m\\k\neq m}} X_{0,k,m} \uu b_k^\dagger\uu b_m\label{phi3}\\
\vec S_{\mathrm{3PC}}\cdot\vec\sigma &=& 2\gamma \sum_{\substack{k,m\\k\neq m}} X_{0,k,m} \la\uu b_m\uu b_k^\dagger-\frac{1}{2}\uu b_k^\dagger\uu b_m\mathbf I\ra, \label{S3}\eea
where the `real-valuedness' of $\varphi_{\mathrm{3PC}}$ and $\vec S_{\mathrm{3PC}}$ follows from the fact that $X_{0,k,m} = X_{0,m,k}^*$. The variance of the phase-shift and of the polarization-rotation angle caused by three-pulse collisions ($k\neq m$) are independent of modulation format since $\uu b_k$ and $\uu b_m$ are statistically independent.
The combined perturbation that is caused by both of the above processes satisfies $\uu a_0 + \Delta \uu a_0^{(\mathrm{2PC})} + \Delta \uu a_0^{(\mathrm{3PC})} = \exp(i\varphi) \exp(\frac{i}{2}\vec S\cdot\vec{\sigma})\uu a_0$, with $\varphi = \varphi_{\mathrm{2PC}}+\varphi_{\mathrm{3PC}}$ and $\vec S = \vec S_{\mathrm{2PC}}+\vec S_{\mathrm{3PC}}$. The unitarity of these processes implies that $\|\uu a_0 + \Delta \uu a^{(\mathrm{2PC})}_{0}+\Delta \uu a^{(\mathrm{3PC})}_{0}\| = \|\uu a_0\|$.

%The effects of four-pulse collisions and of three-pulse collisions of the type $h\neq 0$, $k=m$ remains unchanged relative to the single-polarization case as they produce complex circular noise on each of the two polarizations.

Table 1 summarizes the classification of pulse collisions in the polarization multiplexed case, while pointing out the main properties of each collision type. As seen in the table, the effects of four-pulse collisions and of three-pulse collisions of the type $h\neq 0$, $k=m$ remains unchanged relative to the single-polarization case as they produce complex circular noise on each of the two polarizations.

It is important to note that the effect of nonlinear polarization-rotation is only unitary when both polarization components are considered jointly as a vector. In systems that process the individual polarizations separately, the effect of polarization-rotation does not appear unitary. Rather, from the standpoint of an individual polarization component, the diagonal terms of the polarization-rotation matrix $\exp(\frac{i}{2}\vec S\cdot\vec{\sigma})$ appear as additional phase-noise (which has opposite signs for each of the individual polarization components). Whereas, the non-diagonal terms cause mixing between the two polarization components, which under the assumption that the components are statistically independent, manifests itself as complex circular noise. Consequently, in polarization multiplexed systems where the two polarization channels are processed separately from each other, two-pulse and three-pulse collisions with $h=0$, generate not only pure phase-noise (as in the single-polarization case), but some complex circular noise as well. This implies that the relative significance of phase-noise in such systems is slightly smaller than in the single polarization case. %Figures \ref{Figure3} and \ref{scalingWithM}, as well as all the figures in the section that follows are plotted for the polarization-multiplexed case.

As explained in Secs. \ref{TwoPulse} and \ref{ThreePulse}, and as is summarized in Table 1, the dependence of the NLIN power on modulation format is caused only by the interference terms proportional to $X_{h,m,m}$ (whether $h=0$, or not)\footnote{Notice that the variance of the interference contribution that is proportional to $X_{h,m,m}$ depends on the statistical average of $\|\uu b_m\|^4 = |b_m^{(y)}|^4+2|b_m^{(y)}|^2|b_m^{(x)}|^2+|b_m^{(x)}|^4$, where $b_m^{(y)}$ and $b_m^{(x)}$ are the $y$ and $x$ components of $\uu b_m$, accounting for a dependence on the \emph{fourth-order moment} of the transmitted symbols and for a dependence on the \emph{second-order correlation} between the two polarization components.}. In the polarization multiplexed case, the modulation format dependence of these terms may be slightly weaker than in the case of single polarization. In order to demonstrate this, we examine the NLIN contribution to the $y$ component of $\uu a_0$ that is proportional to $X_{h,m,m}$. This contribution is readily extracted from Eq. (\ref{eq60P}), and it is given by
\bea
i\gamma X_{h,m,m}\hspace{-0.05cm}\la\hspace{-0.00cm} 2 a_h^{(y)}|b_m^{(y)}|^2+a_h^{(y)}|b_m^{(x)}|^2+a_h^{(x)}b_m^{(x)^*}\hspace{-0.05cm}b_m^{(y)}\hspace{-0.05cm}\ra\hspace{-0.07cm}. %a_h^{(y)}b_k^{(y)^*}\hspace{-0.05cm}b_m^{(y)}\hspace{-0.1cm}+a_h^{(y)}b_k^{(x)^*}\hspace{-0.05cm}b_m^{(x)}\hspace{-0.1cm}+a_h^{(x)}b_k^{(x)^*}\hspace{-0.05cm}b_m^{(y)}\hspace{-0.05cm}\ra\hspace{-0.07cm}.
\label{eqPol}
\eea
The first term in Eq. \eqref{eqPol} falls into the category of single polarization interactions and its contribution to the variance of the NLIN is modulation-format dependent, exactly as discussed in Secs. \ref{TwoPulse}-\ref{ThreePulse} for the single-polarization case. The same is true for the second term in Eq. (\ref{eqPol}), which  represents a classical XPM interaction between the $y$-polarized symbols from the channel of interest and the $x$-polarized symbols from the interfering channel. But the third term in Eq. \eqref{eqPol} is different in the sense that it involves interference between four \textit{different} data symbols. Assuming that the data transmitted on different polarizations is statistically independent, the variance of the third term is independent of modulation format, and therefore it does not contribute to the modulation format dependence of the overall NLIN variance. It should be noted however, that the modification to the significance of both phase-noise (when the polarization channels are processed separately) and modulation-format dependence in polarization multiplexed systems is relatively small, as can be seen in \cite{DarJLTInv} and in the section that follows.

%We further address the dependence on modulation format of the phase-noise and polarization-rotation components of the NLIN. Clearly, this dependence is attributed to the contribution of two-pulse collisions to phase noise $\varphi_{\mathrm{2PC}}$ and to polarization-rotation $\vec S_{\mathrm{2PC}}$. As can be readily deduced from Eq. \eqref{phi2}, the phase-noise $\varphi_{\mathrm{2PC}}$ strongly depends on the modulation format, and in particular, it vanishes in the case of phase-modulated transmission (such as QPSK) where $\|\uu b_m\|^2$ is constant. The dependence of the polarization rotating effect on modulation format can be seen by observing how the orientation of the vector $\vec{B}_m$ is scattered on the Poincar\'{e} sphere. In particular, the isotropy of the orientations of $\vec B_m$ on the Poincar\'{e} sphere increases with increasing QAM-order, whereas with BPSK, or QPSK, all the vectors $\vec{B}_m$ reside in a single plane. Since the polarization-rotation is determined by $\sum_m X_{0,m,m}\vec{B}_m$, the variance of the rotations angle $|\vec S_{\mathrm{2PC}}|$ also depends on modulation format (as shown in the section that follows), but it is always non-zero, even in the case of pure phase-modulation. In order to demonstrate these ideas, we examine the NLIN contribution of the term proportional to $X_{0,m,m}$, to the $y$ component of $\uu a_0$. The contribution to the phase noise $\varphi_{\mathrm{2PC}}$ is $\gamma X_{0,m,m} \tfrac{3}{2}\la |b_m^{(y)}|^2 + |b_m^{(x)}|^2\ra$, which is clearly modulation format dependent.

Finally, as we demonstrate in Appendix \ref{PolRot}, the conclusions of this section do not change when the modulated polarization axes in the two channels are not identical.
In particular, the phase-noise $\varphi$, the angle of polarization-rotation $|\vec S|$, and the NLIN variance in each of the two polarizations, are all invariant to whether the modulated polarization axes in the two WDM channels are identical, or not. A feature that changes somewhat when the modulated polarization axes are different in the two channels, is the relative significance of phase-noise in systems where the two polarization components are processed separately. For further details, the reader is referred to Appendix \ref{PolRot}.

%In addition, we show in Appendix \ref{PolRot} that rotation of the relative polarization between the interacting channels may only affect the fraction of NLIN that can be attributed to phase-noise, but not the other properties of NLIN, including its overall variance.

\section{System implications of collision classification\label{System}}
The nature of NLIN in systems changes depending on whether it is dominated by complete or incomplete collisions. A collision can be safely characterized as complete when the evolution of the individual pulses during the collisions is negligible. This condition can be translated into two requirements. The first is that the dispersive broadening of the pulses during the collision is negligible relative to the pulse width prior to the collisions. This requirement is automatically satisfied when the frequency spacing between the interacting WDM channels is sufficiently larger than the bandwidth of the channels themselves. To see that, we define the collision length, $L_c$, as the length of the section of fiber in which the collision takes place $L_c \simeq \tfrac{\Delta t}{\beta''\Omega}$, where $\Delta t$ is the temporal duration of the pulses immediately prior to the collision. Denoting by $B$ the spectral width of the individual pulses, the temporal broadening that is caused by dispersion during the process of a collision can be approximated by $\beta'' B L_c = \Delta t\tfrac{B}{\Omega}$, and it is certain to become much smaller than $\Delta t$ when $\Omega/B$ is large\footnote{As can be seen in Sec. \ref{numerics}, consistency with the pulse-collisions theory is observed even in the case of a 32 G-baud system with a 50 GHz channel separation, where $\Omega/B$ is only of the order of 1.5.}.

\begin{figure}[t]
\center
\includegraphics[width=0.425\textwidth]{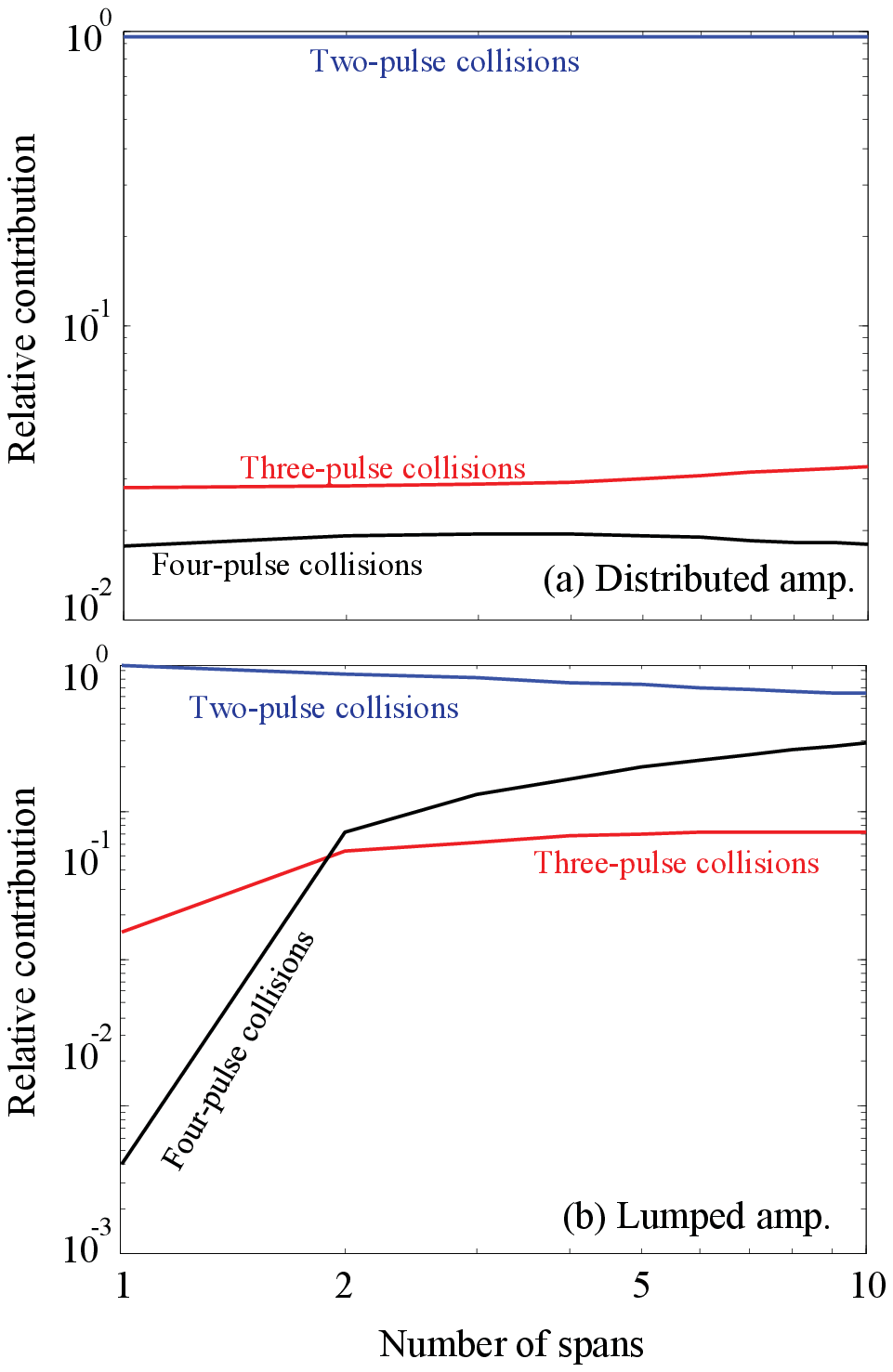}
%\vspace{-0.4cm}
\caption{The contribution of two-pulse collisions $\sum_m \abs{X_{0,m,m}}^2$, three-pulse collisions $\la \sum_{k\neq m} \abs{X_{0,k,m}}^2 + \sum_{h\neq 0} \abs{X_{h,m,m}}^2\ra$, and four-pulse collisions $\sum_{h\neq 0, k\neq m} \abs{X_{h,k,m}}^2$, normalized to the overall NLIN contributions $\sum_{h,k,m} \abs{X_{h,k,m}}^2$, {assuming polarization-multiplexed transmission of 87 WDM channels, at $32$GHz baud-rate, positioned on a $50$GHz ITU grid (43 channels on each side of the channel of interest).} Span length is $100$km where top-panel (a) corresponds to ideally distributed amplification and bottom panel (b) corresponds to lumped amplification.}\label{relCont}
\end{figure}

The second requirement is that the effects of attenuation or gain are negligible during a collision. In systems using lumped amplification this condition is not satisfied when the collision overlaps with an amplifier site, or when the collision length $L_c$ is so long that attenuation during the collision cannot be neglected. In modern uncompensated fiber links, where $\Delta t$ increases due to the accumulated dispersion, $L_c$ is bound to reach values over which fiber loss becomes significant and the collision can no longer be considered complete. For this reason, truly complete collisions may exist in distributed amplification systems (as can be approximately achieved with Raman technology). In systems using lumped amplification, complete collisions occur mostly in the first few spans, where the pulses are still narrow enough to ensure that the collision length is short so that the attenuation during the collision is negligible and that only a small fraction of the collisions overlap with an amplification site.

As can be concluded from the combinations of indices, the number of four-pulse collisions is the largest, whereas the smallest number of collisions are those involving only two pulses. Nevertheless, in a regime where complete collisions dominate, the nature of NLIN is dictated primarily by two-pulse collision processes, whose scaling with frequency separation is the most favorable (proportional to $\Omega^{-1}$). In this regime the NLIN is expected to have a strong phase and polarization-rotation noise character, and its variance should be strongly dependent on the modulation format. In the opposite case, where incomplete collisions dominate, the larger number of four-pulse collisions emphasizes their significance and makes the overall NLIN more similar to complex circular noise. {In order to illustrate these principles we consider in what follows the cases of systems with distributed amplification (governed by complete collisions) and with lumped amplification (where short links are governed by complete collisions and long links are governed by incomplete collisions). We assume a polarization multiplexed transmission of square-root raised-cosine pulses with a roll-off factor of 0.2, propagating over $100$ km spans of standard single-mode fiber ($\beta''=21$ps$^2$/km, $\gamma=1.3$ W$^{-1}$km$^{-1}$, {and attenuation of 0.2 dB/km}). We further assume that the entire C-band (35 nm) is populated with $32$GBaud WDM channels positioned on a $50$GHz ITU grid (amounting to a total of $87$ WDM channels), where the channel of interest is at the center. The choice of using 32GBaud transmission was made for consistency with current commercial systems. The principles governing NLIN build-up as described in this paper will not change at higher symbol-rate. In fact, a shorter collision length will make the contribution of complete collisions even closer to the theoretical prediction. Additionally, in order to facilitate the comparison between the various cases, the examples that we show in this section are plotted for the case of 10-span links regardless of the modulation format, or type of amplification. However, the principles of NLIN build-up do not change when the link extends over longer distances. }

We first examine the various pulse-collision contributions to the quantity $\sum_{h,k,m} \abs{X_{h,k,m}}^2$, which in the case of Gaussian modulation is proportional to the overall NLIN variance (see appendix in \cite{DarJLTInv}). Other modulation formats are considered subsequently. The relative contributions of two-pulse collisions ($h=0$, $k=m$), three-pulse collisions\footnote{{The two types of three-pulse collisions (those with $h=0$ and those with $h\neq 0$) contribute similarly to the overall NLIN variance in the case of Gaussian modulation since $X_{h,m,m}=X_{0,m-h,m}^*$.}}  ($h=0$, $k\neq m$, or $h\neq 0$, $k=m$), and four-pulse collisions ($h\neq 0$, $k\neq m$) to this quantity are plotted in Fig. \ref{relCont} for the cases of distributed and lumped amplification. The dominance of two-pulse collisions is clearly evident in the distributed amplification case, whereas in the case of lumped amplification two-pulse collisions are dominant primarily in few span systems (where a large fraction of the collisions are complete). In longer lumped amplification systems the role of four-pulse collisions is emphasized, as the number of incomplete collisions grows with the number of spans.

\begin{figure}[t]
\center
\includegraphics[width=0.425\textwidth]{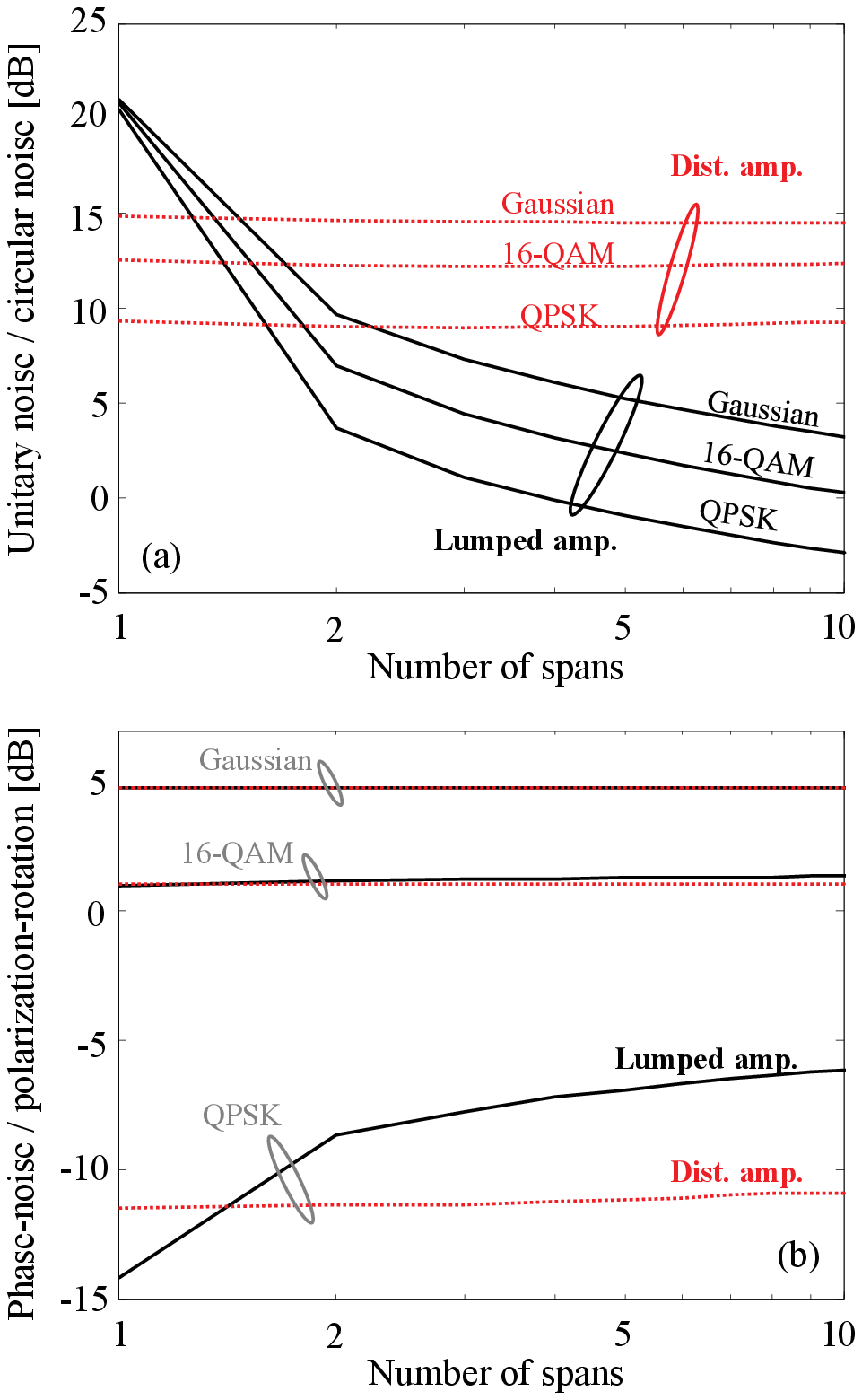}
%\vspace{-0.4cm}
\caption{(a) The ratio between the contribution of unitary noise (phase-noise and polarization-rotation) and circular noise to the overall NLIN power, i.e. $PT(\sigma_\phi^2 +\sigma_S^2)/\sigma_\nu^2$ in the notation of Eq. \eqref{phn}. (b) The ratio between the contributions of phase-noise and polarization-rotation to the overall NLIN power, i.e. $\sigma_\phi^2/\sigma_S^2$. Dotted (red) curves correspond to ideally distributed amplification and solid (black) curves correspond to lumped amplification, {assuming 43 WDM channels on each side of the channel of interest.} }\label{Figure4}
\end{figure}

In order to distinguish between complex circular noise on the one hand, and phase and polarization noise on the other, we express the received {vector of} symbols {at the $n$-th time-slot} as
\bea \uu a_n + \Delta \uu a_n = e^{i\phi_n}e^{\frac i 2 \vec S_n\cdot\vec\sigma}\uu a_n+\uu\nu_n \label{recVec}\eea
where $\phi_n$ represents the phase-noise component of the NLIN, $\vec S_n$ is the polarization-rotation vector, as explained in the previous section, and  $\uu\nu_n$ represents circular nonlinear noise.
%We absorb the average (time independent) nonlinear phase offset into $a_0$ so that both $\phi$ and $n$ are zero mean random variables.
In the relevant limit of signal to noise ratio being much larger than unity $\Delta \uu a_n\simeq i (\phi_n\mathbf I + \frac 1 2 \vec S_n\cdot\vec\sigma)\uu a_n+\uu\nu_n$ and assuming that the data vectors $\uu a_j$ and $\uu b_j$ are statistically independent, identically distributed and isotropically symmetric in the phase-space, the NLIN variance can be expressed as%\footnote{To arrive at Eq. \ref{phn} one needs to demonstrate that $\phi_n\uu a_n$, $(\vec S_n\cdot\vec\sigma)\uu a_n$ and $\nu_n$ are uncorrelated. This can be deduced from Appendix \ref{PolRot}.}
\bea  \sigma_{\mathrm{NLIN}}^2 &=&\lip\|\Delta \uu a_n\|^2\rip - \|\lip\Delta \uu a_n\rip\|^2 \nonumber\\
&=& PT(\sigma_\phi^2 +\sigma_S^2)+ \sigma_\nu^2,\label{phn}\eea
where $\sigma_\phi^2 = \lip\phi_n^2\rip-\lip\phi_n\rip^2$, $\sigma_S^2 = \tfrac{1}{4}\lip|\vec S_n|^2\rip$, $\sigma_\nu^2=\lip\|\uu\nu_n\|^2\rip$, $P$ is the average signal power and $T$ is the symbol duration (so that $\lip\|\uu a_n\|^2\rip=PT$).
The first two terms on the right-hand-side of the bottom row of Eq. \eqref{phn} are the phase-noise and polarization-rotation contributions to the total NLIN variance, whereas the third term represents the contribution of the complex circular noise.

In Fig. \ref{Figure4}a we plot the ratio between the contributions of the unitary noise (i.e., phase-noise and polarization-rotation) and of the circular noise to the overall NLIN variance (namely, $PT(\sigma_\phi^2 +\sigma_S^2)/\sigma_\nu^2$). Notice that this ratio is power independent as all the contributions to the NLIN variance are proportional to $P^3$. In the distributed amplification case, where complete collisions govern the generation of NLIN so that the effect of two-pulse collisions is dominant, phase-noise and polarization-rotation constitute the largest NLIN contribution, as expected. In the case of lumped amplification, where complete collisions dominate only the beginning of the link, phase-noise and polarization-rotation are largest after the first amplified span. As the number of spans increases, collisions become increasingly incomplete and therefore circular noise caused by four-pulse collisions grows gradually.

\begin{figure}[t]
\center
\includegraphics[width=0.425\textwidth]{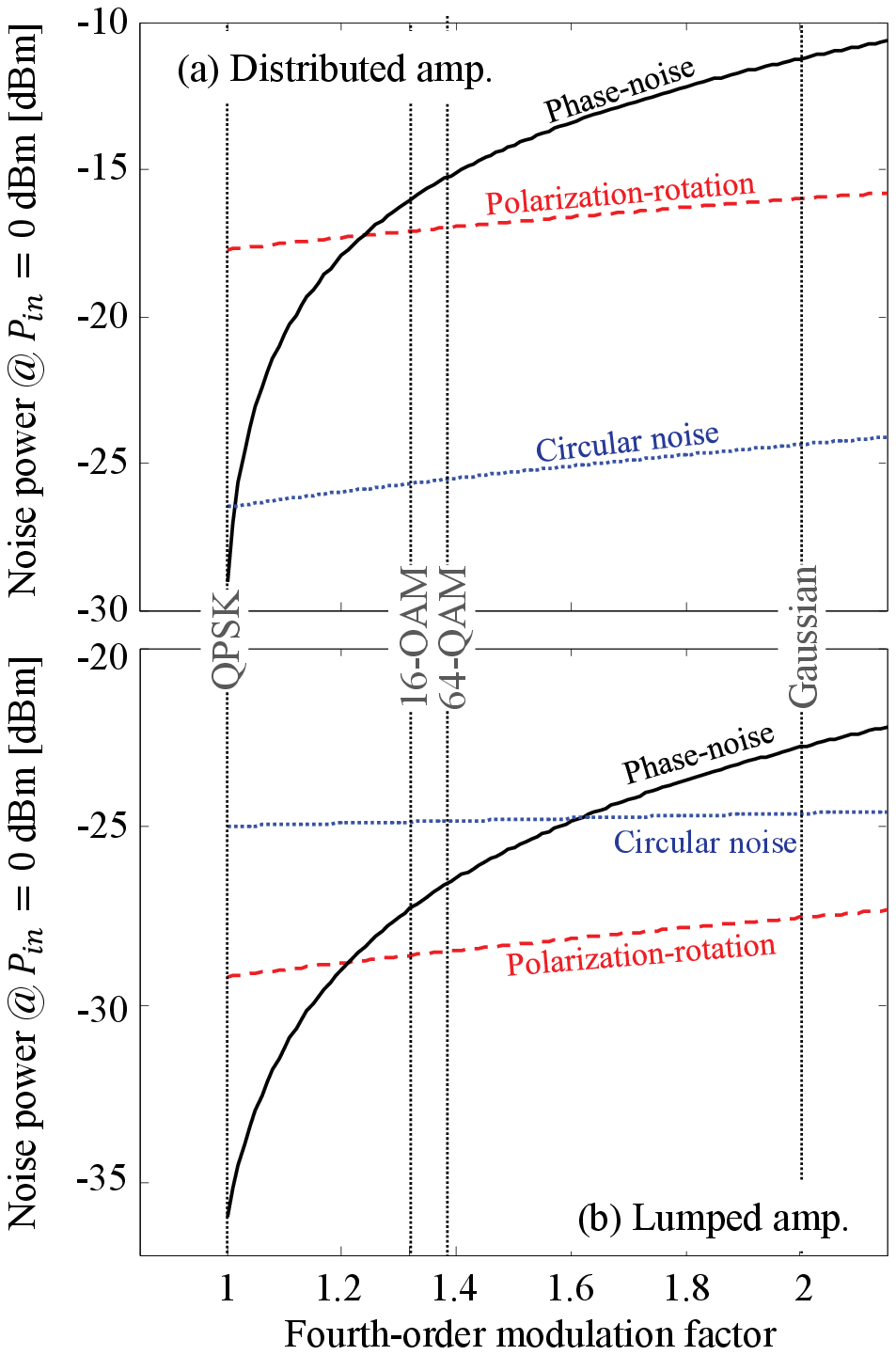}
%\vspace{-0.4cm}
\caption{The contributions of phase-noise (solid black curves), polarization-rotation (dashed red curves), and circular noise (dotted blue curves) to the NLIN power, as a function of the modulation format factor $M${, for a $10\times100$km link, assuming polarization-multiplexed transmission with 43 WDM channels on each side of the channel of interest, baud rate of $32$GHz and $50$GHz channel spacing. Ideally distributed amplification is considered in figure (a) and lumped amplification is considered in figure (b). Values of $M$ corresponding to known modulation formats are marked on the figure.}}\label{Figure5}
\end{figure}

Figure \ref{Figure4}b illustrates the relative significance of the two unitary noise contributions, where the ratio $\sigma_\phi^2/\sigma_S^2$ is plotted versus the number of spans. Clearly, the role of phase-noise increases with the extent to which the intensity of the symbols is modulated, growing from QPSK to 16-QAM and then to Gaussian modulation. This point is further explored in Fig. \ref{Figure5}.
In addition, in the cases of Gaussian modulation and 16-QAM there is almost no difference between the cases of lumped and distributed amplification, whereas a notable difference exists in the case of QPSK. The reason is that in QPSK, only three-pulse collisions (of the type $h=0$) generate phase-noise and their relative significance increases with the number of spans in the case of lumped amplification, whereas it is independent of the link-length in the case of distributed amplification (see Fig. \ref{relCont}).

We further address the dependence of the unitary noise and the circular noise on modulation format. In order to do that we introduce the concept of the {\it Fourth-order modulation factor}
\bea M=\frac{\lip |b|^4\rip}{\langle |b|^2\rangle^2} \label{4modpar}\eea
where the random variable $b$ is a symbol in one polarization component of the interfering channel. The fourth-order modulation factor is equal to unity in the case of pure phase-modulation and grows as the power variations of the modulated signal increase. For example in the case of QAM modulation, $M=1$ for pure phase modulation (like QPSK) and increases asymptotically towards 1.4 with the QAM order (specific values are $M=1.32$ for 16-QAM, and $M=1.38$ for 64-QAM). In the case of complex Gaussian modulation $M$ is equal to $2$. Excluding cases of extremely tight channel-spacing, which require the correction term discussed in \cite{PogEGN}, {and assuming that the two polarization components of the transmitted data vectors are statistically independent,} all aspects of the modulation-format dependence of the NLIN power have been shown to be captured by the fourth-order modulation factor \cite{DarOpex,DarOpex2}.

In Fig. \ref{Figure5} we plot the contributions of phase-noise $PT\sigma_\phi^2$, polarization-rotation $PT\sigma_S^2$, and circular noise $\sigma_\nu^2$ to the NLIN power as a function of $M$ in a $10\times100$km link with distributed (Fig. \ref{Figure5}a) and lumped (Fig. \ref{Figure5}b) amplification. Since the NLIN power scales with $P^3$, the curves are displayed for the case in which $P=1$ mW, and therefore to obtain the results for other input powers, the vertical axes simply need to be shifted upward by three times the average input power expressed in dBm units.
%In the case of distributed amplification it is evident that the phase-noise contribution becomes dominant almost as soon as some level of variation is introduced into the energies of the transmitted symbols. In the case of lumped amplification phase-noise is only dominant in the case of a single-span, whereas in the case of a 5-span system it exceeds the contribution of the circular noise only when $M$ is larger than $\sim1.7$. Note that as we discussed in Secs. \ref{TwoPulse}--\ref{FourPulse}, of all pulse collisions that contribute to circular noise, only the three-pulse collisions that are represented by coefficients of the form $X_{h,m,m}$ introduce modulation format dependence. In the case of mostly complete collisions (distributed amplification), three pulse collisions are stronger than four pulse collisions and therefore the dependence on modulation format (the slope of the dashed-blue curves) is visible in Fig. \ref{Figure5}a. In the case of lumped amplification with more than one fiber span, incomplete four-pulse collisions dominate the generation of circularly symmetric noise and therefore the dependence on modulation format in Fig. \ref{Figure5}b is notably weaker.
Evidently, the contribution of phase-noise strongly depends on modulation format in both cases of distributed and lumped amplification, whereas the dependence of circular noise and polarization-rotation on modulation format is notably weaker. The dependence of the phase-noise on modulation format can be attributed to the fact that it is dominated by two-pulse collisions which, as we have explained earlier, are strongly dependent on modulation format. The circular noise, on the other hand, is strongly affected by four-pulse collisions (particularly in the multi-span lumped amplification scenario), and therefore the dependence of its variance on modulation format is weak. As to polarization-rotations noise, which is also dominated by two-pulse collisions, its weak dependence on modulation format can be understood in view of Eq. \eqref{S2}, which shows that the rotation angle (given by $|\vec S_{\mathrm{2PC}}|$) is determined by a summation that includes the random vectors $\vec B_m$, resulting in a substantial rotation angle variance even in the case of QPSK modulation.

\section{Relation to the Gaussian noise model and the role of chromatic dispersion\label{GN}}
The GN model has been derived under the assumption that the electric fields of the interfering channels are statistically independent Gaussian random processes \cite{Poggiolini,Carena,Pog2,Johannisson,Bononi}. Hence its predictions regarding the NLIN power are rigorously accurate (within the obvious limits of the first-order perturbation analysis) only when the interfering channels undergo Gaussian modulation. With other modulation formats, the GN model is known to have certain inaccuracies, which are particularly conspicuous in the case of distributed amplification systems \cite{DarOpex}, or few-span systems (less than 1000 km) using lumped amplification \cite{Carena,DarOpex2,DarJLTInv}. The improvement in the GN model's accuracy with the number of spans is often attributed to chromatic dispersion \cite{Carena,PogECOC,SerenaECOC13}, with the main argument being that with large chromatic dispersion the electric fields of the propagating channels approach Gaussian statistics and hence the assumption of Gaussianity, on which the GN model relies, is better satisfied. The difficulty with this argument is that it does not explain the fact that in systems using distributed amplification the inaccuracy of the GN model with non-Gaussian modulation does not seem to reduce with the length of the system \cite{DarOpex}, regardless of the amount of chromatic dispersion that the signals accumulate. The role of chromatic dispersion and its relation to the GN model's accuracy can be explained in terms of the pulse collision picture introduced in this paper, as we elaborate in what follows.
 %examined qualitatively in what follows.  The practical problem of the GN model's inaccuracy in predicting the NLIN variance has been resolved in \cite{DarOpex,PogEGN} by performing an analysis that does not assume the Gaussianity of the interferers. But the explanation of the nonlinear interference dynamics and the role that chromatic dispersion plays in it and the relation to the GN model is resolved here.

The GN model is accurate in the regime where modulation-format independent three and four-pulse collisions dominate the formation of NLIN. This does not occur in systems with distributed amplification, where the completeness of collisions implies that the effects of three and four-pulse collisions practically vanishes and the NLIN is dominated by two-pulse collision processes, independently of the link's length, as can be seen in Fig. \ref{relCont}(a). In the regime of lumped optical amplification incomplete collisions become significant to the formation of NLIN only after the first few spans (see Fig. \ref{relCont}(b)), which is where the GN model's accuracy begins to improve.

Another interesting aspect of the role of chromatic dispersion has been introduced in \cite{PogECOC} and followed in \cite{DarOpex2}. It has been shown that in the presence of large pre-dispersion the predictions of the GN model become accurate already in the first span, where in the absence of pre-dispersion the GN model is known to be highly inaccurate. It has been suggested \cite{PogECOC} that the better accuracy of the GN model in this case results from the fact that pre-dispersion transforms the distribution of the electric field of the interfering channels into Gaussian, independently of their modulation format. While tempting in its simplicity, this argument is inconsistent with an observation that was reported in \cite{DarOpex2}, where it was shown that the inaccuracy of the predictions of the GN model in pre-dispersed systems is smaller than without pre-dispersion only in short links whose accumulated dispersion is much smaller than the pre-dispersion that was applied. For example, when the interfering channels are pre-dispersed by an amount equivalent to 500 km of transmission fiber, the GN model's prediction improved in the first few hundred kilometers, but in longer links, of the order of 1000 km, it deteriorated to the same level of inaccuracy characterizing systems without pre-dispersion, as illustrated in Fig. 5 of \cite{DarOpex2}. Clearly, if the GN model's accuracy improved because of better Gaussianity that followed from pre-dispersion, there is no reason for it to deteriorate in longer systems, where even more dispersion is accumulated.
The theory presented in this paper suggests a different explanation, that was briefly alluded to in \cite{DarOpex2}. Large pre-dispersion creates overlaps between all pulses at the link input so that the collisions taking place in the first few spans of the systems are incomplete. The incompleteness of the collisions emphasizes the significance of modulation format independent three and four-pulse collisions whose effect tends to agree with the predictions of the GN model. As the system becomes longer, the impact of pre-dispersion reduces and the inaccuracy of the GN model resumes its regular value.

\section{Conclusions\label{conc}}
We have shown that the various contributions to {inter-channel} NLIN can be classified as pulse collisions involving two, three, or four colliding pulses. Furthermore, we have argued that pulse collisions may either be complete (if they occur on a length-scale where loss/gain and dispersive broadening are negligible), or incomplete. When the majority of collisions are complete (as occurs with distributed gain systems \cite{DarOpex} or in few-span systems using lumped amplification \cite{DarOpex2}) the NLIN is dominated by two-pulse collisions which give it the character of phase-noise {and polarization-rotation noise} with strong dependence on modulation format. When the majority of collisions are incomplete (high span-count systems with lumped amplification), the role of four-pulse interactions becomes more prominent, emphasizing the circular NLIN component, as assumed in \cite{Poggiolini,Carena,Pog2,Johannisson,Bononi}. In this case the dependence of NLIN on the modulation format is much weaker.

\section{Acknowledgement}
The authors would like to acknowledge financial support from the Israel Science Foundation (grant 737/12). Ronen Dar would like to acknowledge the support of the Adams Fellowship of the Israel Academy of Sciences and Humanities, the Yitzhak and Chaya Weinstein Research Institute and the Advanced Communication Center and the Electro-Optic Fund of Tel-Aviv University.

\section{Appendix: The dependence on $\Omega$ in complete versus incomplete collisions}\label{Collisions}
%\subsection{Derivation of $X_{h,k,m}$}
We consider a generic setup of the kind plotted in Fig. \ref{Figure1}. Two pulses, with indices $0$ and $h$, belonging to the channel of interest collide with two pulses of the interfering channel, whose indices are $k$ and $m$. Recall that the pulse with the index $0$ is not a real pulse that propagates along the fiber, but rather (as we described earlier) it is included in the analysis so as to account for the effect of matched filtering.

We assume the regime of large accumulated dispersion, which is relevant in the case of dispersion uncompensated links and take advantage of the far-field approximation \cite{DarOpex,Mecozzi},
\bea
g(z,t)\simeq \sqrt{\frac{i}{2\pi\abs{\beta''}z}}\exp{\la -\frac{it^2}{2\beta''z}\ra} \tilde g\la\frac{t}{\beta''z}\ra, \label{fmdaw}
\eea
where $\tilde g(\omega)=\int g(t)\exp(i\omega t)\df t$ is the Fourier transform of the fundamental waveform $g(t)$. By substituting Eq. \eqref{fmdaw} into Eq. \eqref{eq80} and using the change of variables $\omega = t/\beta'' z$, we get
\bea
\hspace{-0.3cm}X_{h,k,m}\hspace{-0.3cm}&\simeq& \hspace{-0.3cm}\frac{e^{i\Omega hT}}{2\pi\abs{\beta''}} \int_{z_0}^{z_1} \hspace{-0.0cm} \frac{f(z)}{z} \tilde\psi\la \Omega+\tfrac{\bar mT}{\beta''z}\ra e^{i\tfrac{\bar mT}{\beta''z} hT} \df z ,\label{eqA50}
\eea
where $\bar m$ stands for $(m+k-h)/2$ and
\bea
\tilde\psi(u) \hspace{-0.3cm}&=&\hspace{-0.3cm} \frac{1}{2\pi}\int_{-\infty}^\infty \hspace{-0.0cm}\tilde g^*\hspace{-0.0cm}\la \omega+ u+\tfrac{hT}{2\beta''z}\ra \tilde g\hspace{-0.0cm}\la \omega+ u-\tfrac{hT}{2\beta''z}\ra \nonumber\\
&&\hspace{0.6cm} \times \hspace{0.1cm}\tilde g^*\hspace{-0.0cm}\la\omega+\hspace{-0.0cm}\tfrac{(m-k)T}{2\beta''z}\ra\tilde g\hspace{-0.0cm}\la\omega-\hspace{-0.0cm}\tfrac{(m-k)T}{2\beta''z}\ra \nonumber\\
 &&\hspace{0.6cm}\times \hspace{0.1cm} e^{i\omega (h-k+m)T}\df \omega.\hspace{-0.2cm}\label{eqA59}
\eea
Note that the function $\tilde\psi(u)$ reduces to zero when $z$ satisfies $\tfrac{\abs{h}T}{2\abs{\beta''}z}>\tfrac{B}{2}$ or $\tfrac{\abs{m-k}T}{2\abs{\beta''}z}>\tfrac{B}{2}$, where $B$ is the spectral width of the fundamental pulse $g(t)$. In the limit of large dispersion, we can assume that $\frac{\abs{m-k}T}{2\abs{\beta''}z}\ll \tfrac{B}{2}$ so that the product $\tilde g^*\hspace{-0.0cm}\la\omega+\hspace{-0.0cm}\tfrac{(m-k)T}{2\beta''z}\ra\tilde g\hspace{-0.0cm}\la\omega-\hspace{-0.0cm}\tfrac{(m-k)T}{2\beta''z}\ra$ is approximately equal to $\abs{\tilde g\la\omega\ra}^2$. In the same manner, by assuming $\frac{\abs{h}T}{2\abs{\beta''}z}\ll \frac{B}{2}$ we get
\bea
\tilde\psi(u) \simeq \frac{1}{2\pi}\int_{-\infty}^\infty \hspace{-0.2cm}\abs{\tilde g\la \omega+ u\ra }^2\abs{\tilde g\la\omega\ra}^2 e^{i\omega (h-k+m)T}\df \omega.\hspace{-0.2cm}\label{eqA60}
\eea
Since $\tilde \psi(u)\simeq 0$ when $\abs{u}\leq B$ we can conclude that the lower integration limit $z_0$ in Eq. \eqref{eqA50} is the largest of the three quantities $\tfrac{\abs{h}T}{\abs{\beta''}B}$, $\tfrac{\abs{k-m}T}{\abs{\beta''}B}$, and $\tfrac{\bar mT}{\abs{\beta''}(\Omega+B)}$, whereas the upper limit $z_1$ is the smaller between the link length $L$ and $\tfrac{\bar mT}{|\beta''|(\Omega-B)}$.
These integration limits $z_0$ and $z_1$ can be intuitively understood when considering the nature of the nonlinear interaction, as illustrated in Fig. \ref{Figure1}.
In the general case, two pulses from the channel of interest interact with two pulses of the interfering channel, and the nonlinear interaction takes place while all four pulses overlap with each other temporally. The dispersive broadening of the individual pulses after propagating to point $z$ along the fiber is approximately $|\beta''| Bz$. Within the far field approximation, significant overlap between the zeroth and the $h$-th pulses in the channel of interest is achieved when $|\beta''| Bz\gg|h|T$ and similarly the $k$-th and $m$-th pulses in the interfering channel overlap when $|\beta''| Bz\gg|m-k|T$. In this case, the zeroth and $h$-th pulses form a single block of width $\sim|\beta''| Bz$ centered at $hT/2$, whereas the $k$-th and $m$-th pulses form a block of the same width, centered at $(m+k)T/2-\beta\Omega z$. The overlap between the two blocks starts when $(m+k)T/2-\beta\Omega z - \beta'' Bz/2>hT/2+\beta'' Bz/2$ and ends when $(m+k)T/2-\beta\Omega z +\beta'' Bz/2<hT/2-\beta'' Bz/2$. The combination of these conditions translates into the integration limits indicated above, implying that when $h,$ $k$ and $m$ satisfy $\bar{m}>\max\{\abs{h},\abs{k-m}\}(1+\Omega/B)$, a collision is formed if the condition $\bar{m}<(\Omega+B)\abs{\beta''}L/T$ is being satisfied. The collision is completed before the end of the link if $h$, $k$ and $m$ further satisfy $\bar{m}<(\Omega-B)\abs{\beta''}L/T$.

Finally, defining $u=\Omega+\tfrac{\bar mT}{\beta'' z}$, Eq. \eqref{eqA50} can be rewritten as
%\bea
%\hspace{-0.5cm}X_{h,k,m} \hspace{-0.3cm}&=& \hspace{-0.35cm}\frac{1}{2\pi\abs{\beta''}} \hspace{-0.1cm}\int_{u_0}^{u_1} \hspace{-0.3cm}\frac{1}{\Omega-u}f\left(\tfrac{\bar mT}{|\beta''|(\Omega-u)}\right) e^{-iu hT}\tilde\psi(u) \df u \label{eq2forfig}
%\eea
\be
X_{h,k,m} =\frac{1}{2\pi\abs{\beta''}} \hspace{-0.1cm}\int_{u_0}^{u_1} \hspace{-0.3cm}\frac{1}{\Omega-u}f\left(\tfrac{\bar mT}{|\beta''|(\Omega-u)}\right) e^{-iu hT}\tilde\psi(u) \df u, \label{eq2forfig}
\ee
where $u_0=\Omega+\tfrac{\bar mT}{\beta'' z_1}$ and $u_1=\Omega+\tfrac{\bar mT}{\beta'' z_0}$. This form will be useful in what follows.
\subsection{Complete collisions}
In the case of a complete collision, the integrand in Eq. \eqref{eq2forfig} vanishes at the integration boundaries, which can therefore be replaced by $\pm\infty$. In addition, since in the regime of complete collisions, attenuation and gain during the collision must be assumed negligible, we may replace $f\left(\tfrac{\bar mT}{|\beta''|(\Omega-u)}\right)$ with
$f\left(\tfrac{\bar mT}{|\beta''|\Omega}\right)$ and move it outside of the integral. Having done so, we obtain
\bea
X_{h,k,m} \hspace{-0.3cm}&=& \hspace{-0.3cm}\frac{f\left(\tfrac{\bar mT}{|\beta''|\Omega}\right)}{2\pi\abs{\beta''}\Omega} \sum_{n=0}^\infty\int_{-\infty}^{\infty} \frac {u^n} {\Omega^n} e^{-iu hT}\tilde\psi(u) \df u. \label{eqA70}
\eea
Defining $\psi(v) = \tfrac{1}{2\pi}\int_{-\infty}^{\infty}\tilde\psi(u)\exp(-iuv)\df u$, as the inverse Fourier transform of $\tilde\psi(u)$, Eq. \eqref{eqA70} can be rewritten as
\bea X_{h,k,m} \hspace{-0.3cm}&=& \hspace{-0.3cm} \frac{1}{\abs{\beta''}\Omega} f\left(\tfrac{\bar mT}{|\beta''|\Omega}\right)\sum_{n=0}^\infty\la\frac{i}{\Omega}\ra^n\psi^{(n)}(hT)\label{eqA80},\eea
where $\psi^{(n)}(hT)$ is the $n$-th derivative of $\psi(v)$, evaluated at $v=hT$.
From Eq. \eqref{eqA60} it is evident that
\bea \psi(v) = R^*(v)R\big(v-(h-k+m)T\big),\label{eqA81}\eea
where $R(v) = \int g^*(t)g(v-t)dt$ is the temporal autocorrelation function of the fundamental pulse $g(t)$. The orthogonality  condition of Eq. \eqref{eqA10} implies that $R(v)$ has a maximum  when $v=0$ ($R(0)=1$) and that it vanishes when $v/T$ is any integer other than 0.

It is evident from Eq. \eqref{eqA81} that the only contribution that scales with $\Omega^{-1}$ is the one proportional to $\psi^{(0)}(hT)=\psi(hT)$. Substitution of $v=hT$ in Eq. \eqref{eqA81} reveals that $\psi(hT)=R^*(hT)R\big((k-m)T\big)$ is non-zero only when $h=0$ and $k=m$. This is exactly the case of two-pulse collisions discussed earlier. The next contribution to NLIN, which scales as $\Omega^{-2}$, follows from $\psi^{(1)}(hT) = R'(hT)R\big((k-m)T\big)+R(hT)R'\big((k-m)T\big)$ and it is nonzero only in the case of three-pulse collisions with $h=0$ and $k\neq m$ (one pulse in the channel of interest and two pulses in the interfering channel), or with $h\neq0$ and $k=m$ (two pulses in the channel of interest and a single pulse in the interfering channel). Only the term with $n=2$ in the summation in Eq. \eqref{eqA80}, which scales as $\Omega^{-3}$ is nonzero in the case of complete four-pulse collisions (with $h\neq0$ and $k\neq m$). Hence we may summarize that complete two-pulse collisions scale as $\Omega^{-1}$, complete three-pulse collisions scale as $\Omega^{-2}$ and complete four-pulse collisions scale as $\Omega^{-3}$. As can be deduced from the integration limits of Eq. \eqref{eq2forfig}, the number of complete two, three and four-pulse collisions is proportional to $\Omega$, $\Omega^2$ and $\Omega^3$, respectively, and therefore their overall contribution to NLIN (which is proportional to $\abs{X_{h,k,m}}^2$) is dominated by two-pulse collisions (see Sec. \ref{numerics}).
\subsection{Incomplete collisions}
The analysis of incomplete collisions is somewhat more complicated and difficult to extend beyond what is given by expressions \eqref{eqA50}, or \eqref{eq2forfig}. Some insight can be gained from considering the case of distributed amplification where $f(z)=1$ and can be taken out of the integral. In this case Eq. \eqref{eqA70} still holds, except that the actual integration limits $u_0$ to $u_1$ need to be applied.

Using the formulation of the previous subsection $\psi(v)$ is no longer given by Eq. \eqref{eqA81}, but rather it is the convolution of the expression on the right-hand-side of Eq. \eqref{eqA81} with a sinc function corresponding to the inverse Fourier transform of a unit frequency window extending between $u_0$ and $u_1$. In this situation $\psi(hT)$ may differ from zero for all combinations of $h,k$, and $m$, implying that the contributions of all types of incomplete pulse collisions scale as $\Omega^{-1}$.

\section{Appendix: Relative rotation between the modulated polarization axes of the interacting WDM channels}\label{PolRot}
In the analysis of polarization multiplexing in Sec. \ref{polarization}, we assumed for simplicity that the modulated polarization axes in the interfering channel are parallel to those in the channel of interest. This assumption clearly does not represent the situation in reality where the modulated polarization axes in each of the two channels undergo different paths between the individual transmitters and the point at which they are wavelength multiplexed into a single transmission fiber. In addition, polarization mode dispersion in practical links may also cause relative polarization rotations between the various WDM channels. We examine these situations in this appendix.

Rigorously, the rotation of relative polarizations can be taken into account by replacing the vectors $\uu b_j$ with $ \mathbf U\uu b_j$, where $\mathbf U$ is an arbitrary unitary matrix representing a relative rotation between the channel of interest and the interfering channel, changing the expression for the NLIN into
\bea \Delta \uu a_0 \hspace{-0.3cm}&=&\hspace{-0.3cm} i \gamma \sum_{h,k,m}  X_{h,k,m} \mathbf U\left(\uu b_k^\dagger \uu b_m \mathbf I+ \uu b_m\uu b_k^\dagger\right)\mathbf U^\dagger   \uu a_h  . \label{eq60P_app} \eea
According to Eq. \eqref{eq60P_app}, and by assuming that the components of $\uu a_j$ and $\uu b_j$ are isotropic in their phase-space, the NLIN variance is given by
\bea
\hspace{-0.2cm}\lip\|\Delta \uu a_0\|^2\rip \hspace{-0.3cm}&-&\hspace{-0.3cm} \|\lip\Delta \uu a_0\rip\|^2 \nonumber\\
&&\hspace{-0.9cm}=\gamma^2 \hspace{-0.3cm}\sum_{\substack{h,k,m\\h',k',m'}}  \hspace{-0.3cm}X_{h,k,m}^* X_{h',k',m'}\lip \uu a_h^\dagger\mathbf{\bar{\Lambda}}_{k,m}^\dagger \mathbf{\bar{\Lambda}}_{k',m'} \uu a_{h'}\rip  ,
\eea
where $\mathbf{\Lambda}_{k,m} = \mathbf U\left(\uu{b}_k^\dagger \uu{b}_m \mathbf I+ \uu{b}_m\uu{b}_k^\dagger-\lip\uu{b}_k^\dagger\uu{b}_m \mathbf I+ \uu{b}_m\uu{b}_k^\dagger\rip\right)\mathbf U^\dagger$, and the subtraction of the average $\lip\uu{b}_k^\dagger\uu{b}_m \mathbf I+ \uu{b}_m\uu{b}_k^\dagger\rip$ inside the parenthesis eliminates the average (deterministic) phase shift of the entire constellation\footnote{The subtracted sum $i\gamma\sum_{h,k,m} X_{h,k,m}\mathbf U\lip\uu{b}_k^\dagger\uu{b}_m \mathbf I+ \uu{b}_m\uu{b}_k^\dagger\rip\mathbf U^\dagger\uu a_h$ represents a deterministic phase-shift when $k=m$ and $h=0$. The summation over the terms $k\neq m$ is clearly 0 because of the statistical independence between symbols and their isotropy. The summation over the terms $h\neq 0$ can also be shown to be 0 by using the property that $\sum_m X_{h,m,m}=0$ for all $h\neq 0$.}. By further assuming that the polarization components of $\uu a_j$ and $\uu b_j$ are statistically independent and identically distributed, it can be readily shown that $\lip \uu a_h^\dagger\mathbf{\bar{\Lambda}}_{k,m}^\dagger \mathbf{\bar{\Lambda}}_{k',m'} \uu a_{h'}\rip$ is zero whenever $h'\neq h$, $k'\neq k$, or $m'\neq m$, meaning that the NLIN variance is
\bea
\hspace{-0.2cm}\lip\|\Delta \uu a_0\|^2\rip \hspace{-0.3cm}&-&\hspace{-0.3cm} \|\lip\Delta \uu a_0\rip\|^2 \nonumber\\
&=&\gamma^2 \sum_{h,k,m}  \lip \uu a_h^\dagger\mathbf{\bar{\Lambda}}_{k,m}^\dagger \mathbf{\bar{\Lambda}}_{k,m} \uu a_h\rip |X_{h,k,m}|^2  \nonumber\\
\hspace{-0.3cm}&=&\hspace{-0.3cm} \gamma^2 \sum_{h,k,m}  \mathrm{Tr}\la\lip \mathbf{\bar{\Lambda}}_{k,m} \uu a_h\uu a_h^\dagger\mathbf{\bar{\Lambda}}_{k,m}^\dagger \rip\ra |X_{h,k,m}|^2 \nonumber \\
\hspace{-0.3cm}&=&\hspace{-0.3cm} \gamma^2 \lip|a|^2\rip\sum_{h,k,m}  \lip\mathrm{Tr}\la \mathbf{\bar{\Lambda}}_{k,m}\mathbf{\bar{\Lambda}}_{k,m}^\dagger \ra\rip |X_{h,k,m}|^2 ,\label{eq60P_app2}
\eea
where $\mathrm{Tr}(\mathbf A)$ denotes the \emph{trace} of the matrix $\mathbf A$. Finally, since the trace of a matrix is invariant to unitary rotations, i.e. $\mathrm{Tr}\la \mathbf U\mathbf A\mathbf U^\dagger\ra=\mathrm{Tr}\la \mathbf A \ra$, Eq. \eqref{eq60P_app2}, and therefore also the NLIN variance, are independent of the relative rotation between the channels. Notice that symmetry between the two polarization components suggests that the NLIN variance in each polarization component is also invariant to $\mathbf U$.

Further insight into the effect of relative polarization rotation can be extracted from Eq. \eqref{recVec}, which describes the effect of nonlinearity as a phase-noise $\phi_n$, polarization-rotation noise $\vec S_n$, and circular noise $\nu_n$. The variance of the circular noise $\nu_n$ is independent of $\mathbf U$, as discussed in the previous paragraph. In addition, using Eqs. (\ref{phi2}-\ref{S3}) it can be easily verified that both the phase shift $\phi_n$ and the rotation angle $|\vec S_n|$ are also independent of $\mathbf U$. Yet, in cases where the two polarization channels are processed separately from each other, the inclusion of $\mathbf U$ may affect the phase-noise that is observed in each of the individual polarization components. This can be seen most conveniently from the contribution of two-pulse collisions to the phase-noise in the $y$ component of the vector $\uu a_0$, which are given by
%example given in Sec. \ref{polarization} applied to the case of two-pulse collisions ($h=0$, $k=m$)
%
\bea
\gamma \sum_m X_{0,m,m}\hspace{-0.05cm}\la\hspace{-0.00cm} 2 |\tilde b_m^{(y)}|^2+|\tilde b_m^{(x)}|^2\ra\hspace{-0.07cm}, %a_h^{(y)}b_k^{(y)^*}\hspace{-0.05cm}b_m^{(y)}\hspace{-0.1cm}+a_h^{(y)}b_k^{(x)^*}\hspace{-0.05cm}b_m^{(x)}\hspace{-0.1cm}+a_h^{(x)}b_k^{(x)^*}\hspace{-0.05cm}b_m^{(y)}\hspace{-0.05cm}\ra\hspace{-0.07cm}.
\label{eqPol2}
\eea
where $\tilde b_m^{(x)}$ and $\tilde b_m^{(y)}$ are the $x$ and $y$ component of the vector $\mathbf U\uu b$, respectively. Assume for simplicity the example of phase-modulated transmission (e.g. QPSK) where there is no rotaion of the relative modulation axes ($\mathbf U = \mathbf I$). In this case  $|\tilde b_m^{(x)}|^2=|\tilde b_m^{(y)}|^2=\text{Const}$ and the variance of the phase-noise due to the contribution of two-pulse collisions is 0. In the case where $\mathbf U \neq \mathbf I$, $|\tilde b_m^{(x)}|^2$ and $|\tilde b_m^{(y)}|^2$ become dependent on the transmitted data so that two-pulse collisions contribute to a phase-noise with nonzero variance.

%\vspace{-0.25cm}

%\end{multicols}

\end{document}